\newcounter{myctr}
\def\myitem{\refstepcounter{myctr}\bibfont\noindent\ifnum\themyctr>9\else\phantom{0}\fi\hangindent17pt\themyctr.\enskip}

\documentclass{ws-ijqi}

\usepackage{braket}

\newtheorem{thm}{Theorem}

\begin{document}

\bibliographystyle{unsrt}

\markboth{Takuya Machida}
{Phase transition of an open quantum walk}

\catchline{}{}{}{}{}

\title{Phase transition of an open quantum walk}

\author{Takuya Machida}

\address{College of Industrial Technology, Nihon University, Narashino, Chiba 275-8576, Japan\\
machida.takuya@nihon-u.ac.jp}

\maketitle


\begin{abstract}
It has been discovered that open quantum walks diffusively distribute in space, since they were introduced in 2012.
Indeed, some limit distributions have been demonstrated and most of them are described by Gaussian distributions.
We operate an open quantum walk on $\mathbb{Z}=\left\{0, \pm 1, \pm 2,\ldots\right\}$ with parameterized operations in this paper, and study its 1st and 2nd moments so that we find its standard deviation.
The standard deviation tells us whether the open quantum walker shows diffusive or ballistic behavior, which results in a phase transition of the walker.
\end{abstract}

\keywords{Open quantum walk, Phase transition}

\section{Introduction}
As quantum counterparts of random walks, quantum walks were introduced around 2000 and have been investigated numerically and theoretically~\cite{Gudder1988,AharonovDavidovichZagury1993,Meyer1996,Ambainis2003}.
Their systems are described in Hilbert spaces and the walkers are operated by unitary operations, resulting in different behavior from classical random walks.
The interesting things coming from quantum walks, make a possibility to develop other fields.
Quantum walks have been, for instance, applied to quantum algorithms in quantum computing~\cite{Venegas-Andraca2008}.
While probability distributions of random walks are generally diffusive, the ones of quantum walks are ballistic.
One can confirm the difference in some limit distributions after the walkers have repeated updating their systems a lot of times~\cite{Venegas-Andraca2012}.

In 2012, another type of quantum walk, named  open quantum walk, was introduced~\cite{AttalPetruccioneSabotSinayskiy2012,AttalPetruccioneSinayskiy2012,SinayskiyPetruccione2013}.
Although probability distributions of open quantum walks have not been analyzed enough at this point, specific walks were solved (e.g. ~\cite{SinayskiyPetruccione2012}) and some limit theorems were discovered.
The open quantum walks on $\mathbb{Z}=\left\{0, \pm 1, \pm 2,\ldots\right\}$, which have ever been studied, were not ballistic but diffusive, and it was discovered that their probability distributions converge to a Gaussian distribution or a mixture of Gaussian distributions~\cite{KonnoYoo2013,AttalGuillotin-PlantardSabot2015,CarbonePautrat2015}.
Konno and Yoo~\cite{KonnoYoo2013} demonstrated specific limit distributions, referring a central limit theorem  which had been reported in~\cite{AttalGuillotin-PlantardSabot2015}.

We study an open quantum walk on $\mathbb{Z}$ and the walker launches with a localized initial state.
The operations on the walker are parameterized by two values, that will be represented by $\theta_0$ and $\theta_1$.
The finding probability mostly looks like a Gaussian distribution and diffusively spreads as the walker repeats updating its system.
Some of them, however, seem to have ballistic behavior.
To prove the ballistic behavior, we aim at estimating the standard deviation which will be reported in a limit theorem at the end.     
As a result, we will find a phase transition of the open quantum walk between diffusive and ballistic behavior.

\section{An open quantum walk on $\mathbb{Z}$}

The system of the open quantum walk at time $t\,(=0,1,2,\ldots)$, represented by $D_t$, is defined on the tensor Hilbert space $\mathcal{H}_p\otimes\mathcal{H}_d$.
The position Hilbert space $\mathcal{H}_p$ is spanned by the orthonormal basis $\bigl\{\,\ket{x} : x\in\mathbb{Z}\,\bigr\}$.
Note that the notations $\ket{x}\,(x\in\mathbb{Z})$ could be replaced with the other notations $\ket{x}\bra{x}\,(x\in\mathbb{Z})$ because both notations represent the positions of the open quantum walker.
The Hilbert space $\mathcal{H}_d$ is spanned by the orthonormal basis $\bigl\{\,\ket{0}\bra{0},\, \ket{0}\bra{1},\, \ket{1}\bra{0},\, \ket{1}\bra{1}\,\bigr\}$ where
\begin{equation}
 \ket{0}=\begin{bmatrix}
	  1\\0
	 \end{bmatrix},\quad
 \ket{1}=\begin{bmatrix}
	  0\\1
	 \end{bmatrix}.
\end{equation}

We study the most basic open quantum walk which shifts to the next neighbors. 
The walker moves to the left and to the right with shift operations 
\begin{align}
 S_{-}=& \sum_{x\in\mathbb{Z}}\ket{x-1}\bra{x}\otimes \left(\sum_{j=0}^1\ket{j}\bra{j}\right),\\
 S_{+}=& \sum_{x\in\mathbb{Z}}\ket{x+1}\bra{x}\otimes \left(\sum_{j=0}^1\ket{j}\bra{j}\right),
\end{align}
after the state is changed at each location by local operations 
\begin{align}
 P(\theta_0,\theta_1)
 =& \cos\theta_0\ket{0}\bra{1} + \sin\theta_0\sin\theta_1\ket{1}\bra{0} - \sin\theta_0\cos\theta_1\ket{1}\bra{1}\nonumber\\[1mm]
 =& c_0\ket{0}\bra{1} + s_0s_1\ket{1}\bra{0} - s_0c_1\ket{1}\bra{1}\nonumber\\[1mm]
  =& \begin{bmatrix}
     0 & c_0\\
     s_0s_1 & -s_0c_1
    \end{bmatrix},\label{eq:matrix_P}\\[1mm]
 Q(\theta_0,\theta_1)
 =& \sin\theta_0\cos\theta_1\ket{0}\bra{0} + \sin\theta_0\sin\theta_1\ket{0}\bra{1} + \cos\theta_0\ket{1}\bra{0}\nonumber\\[1mm]
 =& s_0c_1\ket{0}\bra{0} + s_0s_1\ket{0}\bra{1} + c_0\ket{1}\bra{0}\nonumber\\[1mm]
 =& \begin{bmatrix}
     s_0c_1 & s_0s_1\\
     c_0 & 0
    \end{bmatrix},\label{eq:matrix_Q}
\end{align}
where $\theta_0, \theta_1\in [0,2\pi)$ and $c_j=\cos\theta_j,\, s_j=\sin\theta_j\,(j=0,1)$.
The one-step progression of the open quantum walk is, therefore, defined by
\begin{align}
 D_{t+1}
 =& S_{-}\left(\sum_{x\in\mathbb{Z}}\ket{x}\bra{x}\otimes P(\theta_0,\theta_1)\right)D_t\left(\sum_{x\in\mathbb{Z}}\ket{x}\bra{x}\otimes P(\theta_0,\theta_1)^{\ast}\right)\nonumber\\
 & + S_{+}\left(\sum_{x\in\mathbb{Z}}\ket{x}\bra{x}\otimes Q(\theta_0,\theta_1)\right)D_t\left(\sum_{x\in\mathbb{Z}}\ket{x}\bra{x}\otimes Q(\theta_0,\theta_1)^{\ast}\right),\label{eq:time-progress}
\end{align}
where $P(\theta_0,\theta_1)^{\ast}$ and $Q(\theta_0,\theta_1)^{\ast}$ are the adjoint matrices of $P(\theta_0,\theta_1)$ and $Q(\theta_0,\theta_1)$, respectively.
Given an initial state
\begin{align}
 D_0=& \ket{0}\otimes \Bigl(p\ket{0}\bra{0} + (1-p)\ket{1}\bra{1}\Bigr)\nonumber\\
 =& \ket{0}\otimes \begin{bmatrix}
		    p & 0\\
		    0 & 1-p
		   \end{bmatrix}\,\in\mathcal{H}_p\otimes\mathcal{H}_d,
\end{align}
with $p\in [0,1]$, the walker is observed at position $x$ at time $t$ with probability
\begin{equation}
 \mathbb{P}(X_t=x)=\mbox{Tr}\left[\left(\bra{x}\otimes\sum_{j=0}^1\ket{j}\bra{j}\right)D_t\right].
\end{equation}

\section{Phase transition}
We are going to estimate the 1st moment $\mathbb{E}(X_t)=\sum_{x\in\mathbb{Z}} x\,\mathbb{P}(X_t=x)$ and the 2nd moment $\mathbb{E}(X_t^2)=\sum_{x\in\mathbb{Z}} x^2\,\mathbb{P}(X_t=x)$, and see how the standard deviation $\sigma(X_t)=\sqrt{\mathbb{E}(X_t^2)-\mathbb{E}(X_t)^2}$ behaves for large values of time $t$. 

Let us start to define the Fourier transform of the open quantum walk,
\begin{equation}
 \hat\rho_t(k)
 =\sum_{x\in\mathbb{Z}} e^{-ikx}\left(\bra{x}\otimes\sum_{j=0}^1\ket{j}\bra{j}\right)\,D_t\quad (k\in [-\pi, \pi)).
\end{equation}
Note that the notation $i$ represents the imaginary unit, $i=\sqrt{-1}$, in this paper.
The inverse Fourier transform reproduces the system of the open quantum walk,
\begin{equation}
 D_t=\sum_{x\in\mathbb{Z}}\ket{x}\otimes \int_{-\pi}^{\pi}e^{ikx}\,\hat\rho_t(k)\,\frac{dk}{2\pi},
\end{equation}
from which the finding probability is computed,
\begin{align}
 \mathbb{P}(X_t=x)=& \mbox{Tr}\left[\,\int_{-\pi}^{\pi}e^{ikx}\,\hat\rho_t(k)\,\frac{dk}{2\pi}\,\right]\nonumber\\
 =& \int_{-\pi}^{\pi}e^{ikx}\,\Bigl(\bra{0}\hat\rho_t(k)\ket{0}+\bra{1}\hat\rho_t(k)\ket{1}\Bigr)\,\frac{dk}{2\pi}.
\end{align}
The one-step progression of the Fourier transform comes from Eq.~\eqref{eq:time-progress}, 
\begin{equation}
 \hat\rho_{t+1}(k)
 =e^{ik}P(\theta_0,\theta_1)\,\hat\rho_t(k)\,P(\theta_0,\theta_1)^{\ast} + e^{-ik}Q(\theta_0,\theta_1)\,\hat\rho_t(k)\,Q(\theta_0,\theta_1)^{\ast},
\end{equation}
with the initial state $\hat\rho_0(k)=p\ket{0}\bra{0}+(1-p)\ket{1}\bra{1}$.

Here, we rearrange the four components of $\hat\rho_t(k)$ in the form of vector,
\begin{align}
 \ket{\hat\psi_t(k)}
  =& \bra{0}\hat\rho_t(k)\ket{0}\ket{00}
  +\bra{1}\hat\rho_t(k)\ket{1}\ket{01}\nonumber\\[1mm]
  & +\bra{1}\hat\rho_t(k)\ket{0}\ket{10}
  +\bra{0}\hat\rho_t(k)\ket{1}\ket{11}\nonumber\\[1mm]
 =& \begin{bmatrix}
     \bra{0}\hat\rho_t(k)\ket{0}\\
     \bra{1}\hat\rho_t(k)\ket{1}\\
     \bra{1}\hat\rho_t(k)\ket{0}\\
     \bra{0}\hat\rho_t(k)\ket{1}
    \end{bmatrix},
\end{align}
where $\ket{j_1 j_2}=\ket{j_1}\otimes\ket{j_2}\, (j_1, j_2\in\left\{0,1\right\})$, which means
\begin{equation}
 \ket{00}=\begin{bmatrix}
	   1\\0\\0\\0
	  \end{bmatrix},\quad
 \ket{01}=\begin{bmatrix}
	   0\\1\\0\\0
	  \end{bmatrix},\quad
 \ket{10}=\begin{bmatrix}
	   0\\0\\1\\0
	  \end{bmatrix},\quad
 \ket{11}=\begin{bmatrix}
	   0\\0\\0\\1
	  \end{bmatrix}.
\end{equation}
The probability distribution is computed from $\ket{00}$ and $\ket{01}$ components,
\begin{align}
 \mathbb{P}(X_t=x)
 =& \Bigl(\bra{00}+\bra{01}\Bigr)\int_{-\pi}^{\pi}e^{ikx}\,\ket{\hat\psi_t(k)}\,\frac{dk}{2\pi}\nonumber\\
 =& \int_{-\pi}^{\pi}e^{ikx}\,\Bigl(\braket{00|\hat\psi_t(k)} + \braket{01|\hat\psi_t(k)}\Bigr)\,\frac{dk}{2\pi}.
\end{align}

Reproducing the product of $2\times 2$ matrices by the product of $4\times 4$ matrices and a vector,
\begin{align}
 &
 \begin{bmatrix}
  X & Y\\
  Z & W
 \end{bmatrix}
 =
 \begin{bmatrix}
  a & b\\
  c & d
 \end{bmatrix}
 \begin{bmatrix}
  x & y\\
  z & w
 \end{bmatrix}
 \begin{bmatrix}
  \overline{a} & \overline{c}\\
  \overline{b} & \overline{d}
 \end{bmatrix}\nonumber\\[1mm]
 \Leftrightarrow &
 \begin{bmatrix}
  X\\W\\Y\\Z
 \end{bmatrix}
 =
 \begin{bmatrix}
  \overline{a} & 0 & 0 & \overline{b}\\
  0 & \overline{d} & \overline{c} & 0\\
  \overline{c} & 0 & 0 & \overline{d}\\
  0 & \overline{b} & \overline{a} & 0
 \end{bmatrix}
 \begin{bmatrix}
  a & 0 & 0 & b\\
  0 & d & c & 0\\
  c & 0 & 0 & d\\
  0 & b & a & 0
 \end{bmatrix}
 \begin{bmatrix}
  x\\w\\y\\z
 \end{bmatrix},
\end{align}
we find the one-step progression of $\ket{\hat\psi_t(k)}$,
\begin{align}
 & \ket{\hat\psi_{t+1}(k)}\nonumber\\[1mm]
 &= e^{ik}\begin{bmatrix}
	   0 & 0 & 0 & c_0\\
	   0 & -s_0c_1 & s_0s_1 & 0\\
	   s_0s_1 & 0 & 0 & -s_0c_1\\
	   0 & c_0 & 0 & 0
	  \end{bmatrix}^2\ket{\hat\psi_t(k)}
 + e^{-ik}\begin{bmatrix}
	   s_0c_1 & 0 & 0 & s_0s_1\\
	   0 & 0 & c_0 & 0\\
	   c_0 & 0 & 0 & 0\\
	   0 & s_0s_1 & s_0c_1 & 0
	  \end{bmatrix}^2\ket{\hat\psi_t(k)}\nonumber\\[1mm]
 &= \hat{U}(k)\ket{\hat\psi_t(k)},
\end{align}
where
\begin{equation}
 \hat{U}(k)
  =e^{ik}\begin{bmatrix}
	   0 & 0 & 0 & c_0\\
	   0 & -s_0c_1 & s_0s_1 & 0\\
	   s_0s_1 & 0 & 0 & -s_0c_1\\
	   0 & c_0 & 0 & 0
	\end{bmatrix}^2
  + e^{-ik}\begin{bmatrix}
	   s_0c_1 & 0 & 0 & s_0s_1\\
	   0 & 0 & c_0 & 0\\
	   c_0 & 0 & 0 & 0\\
	   0 & s_0s_1 & s_0c_1 & 0
	  \end{bmatrix}^2.
\end{equation}

Let $\lambda_j(k)\,(j=1,2,3,4)$ be the eigenvalues of the matrix $\hat{U}(k)$.
Then, we denote the eigenvector associated to the eigenvalue $\lambda_j(k)$ by $\ket{v_j(k)}$. 
If the set of eigenvectors $\ket{v_j(k)}\,(j=1,2,3,4)$ is linearly independent,
the decomposition of the initial state
\begin{equation}
 \ket{\hat\psi_0(k)}=\sum_{j=1}^4 a_j(k)\ket{v_j(k)},
\end{equation}
drives the vector $\ket{\hat\psi_t(k)}$ into the eigenspace,  
\begin{equation}
 \ket{\hat\psi_t(k)}= \hat{U}(k)^t\ket{\hat\psi_0(k)}
 = \sum_{j=1}^4 \lambda_j(k)^t a_j(k)\ket{v_j(k)}.
\end{equation}
With the notations
\begin{equation}
 w_j(k)=\Bigl(\bra{00}+\bra{01}\Bigr)\ket{v_j(k)}\quad (j=1,2,3,4),
\end{equation}
one can see the representations of the 1st and 2nd moments in the eigenspace,
\begin{align}
 \mathbb{E}(X_t)
 = & \Bigl(\bra{00}+\bra{01}\Bigr)\left(i\frac{d}{dk} \ket{\hat\psi_t(k)}\right)\bigg|_{k=0}\nonumber\\
 = & \sum_{j=1}^4 i\,t\, \lambda_j(0)^{t-1}\lambda'_j(0)a_j(0)w_j(0) + i\,\lambda_j(0)^t\Bigl(a_j(k)w_j(k)\Bigr)'\Big|_{k=0},\\
 \mathbb{E}(X_t^2)
 = & \Bigl(\bra{00}+\bra{01}\Bigr)\left(i^2\frac{d^2}{dk^2} \ket{\hat\psi_t(k)}\right)\bigg|_{k=0}\nonumber\\
 = & \sum_{j=1}^4 -\, t^2\, \lambda_j(0)^{t-2}\left(\lambda'_j(0)\right)^2 a_j(0)w_j(0)\nonumber\\
 & +\,t\,\biggl[\lambda_j(0)^{t-2}\left(\lambda'_j(0)\right)^2 a_j(0)w_j(0)\nonumber\\
 & -\,2\lambda_j(0)^{t-1}\lambda'_j(0)\Bigl(a_j(k)w_j(k)\Bigr)'\Big|_{k=0}\nonumber\\
 & -\,\lambda_j(0)^{t-1}\lambda''_j(0)a_j(0)w_j(0)\biggr]\nonumber\\
 & -\,\lambda_j(0)^t\Bigl(a_j(k)w_j(k)\Bigr)''\Big|_{k=0}.
\end{align}

\clearpage

We are going to compute the 1st and 2nd moments precisely, starting with numerical experiments of probability distributions.
The analysis is demonstrated in some cases. 
\subsection{Case: $c_0=s_0s_1$}

\begin{figure}[h]
\begin{center}
 \begin{minipage}{50mm}
  \begin{center}
   \includegraphics[scale=0.4]{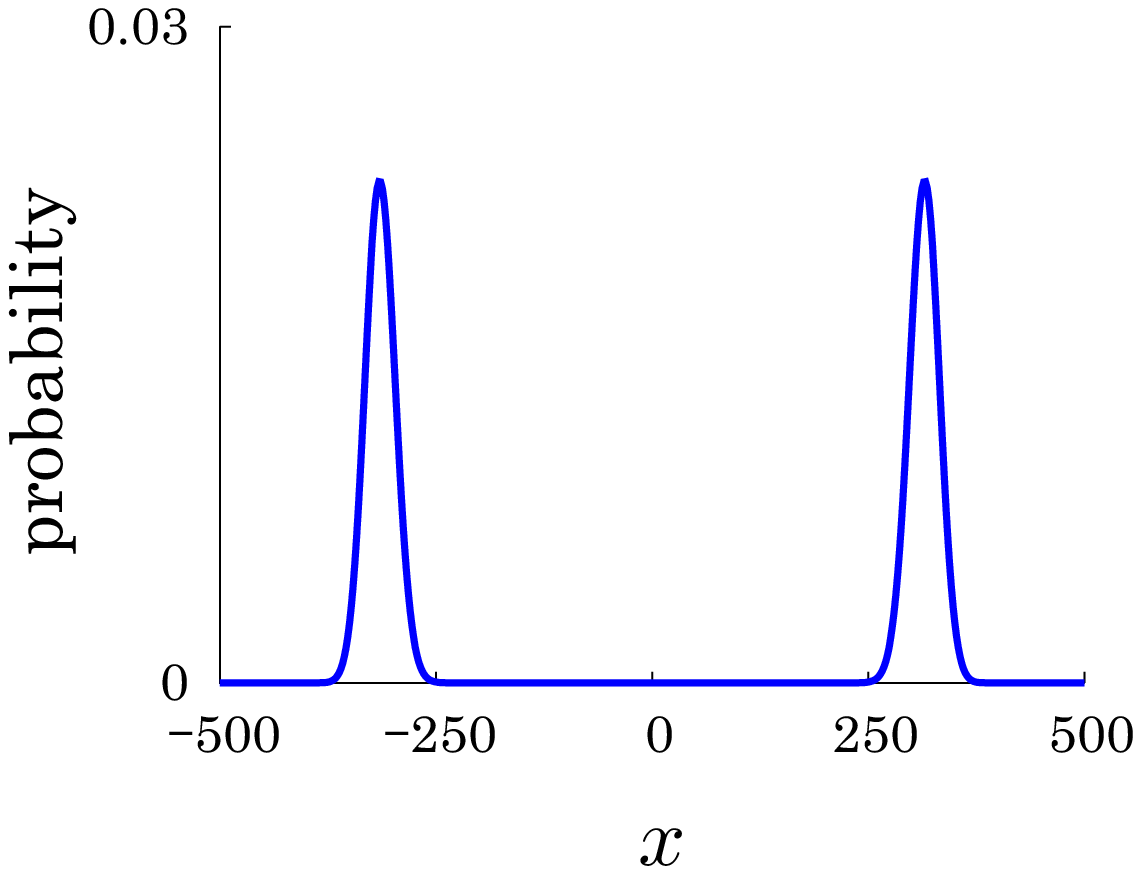}\\[2mm]
  (a) $p=1/2$
  \end{center}
 \end{minipage}
 \begin{minipage}{50mm}
  \begin{center}
   \includegraphics[scale=0.4]{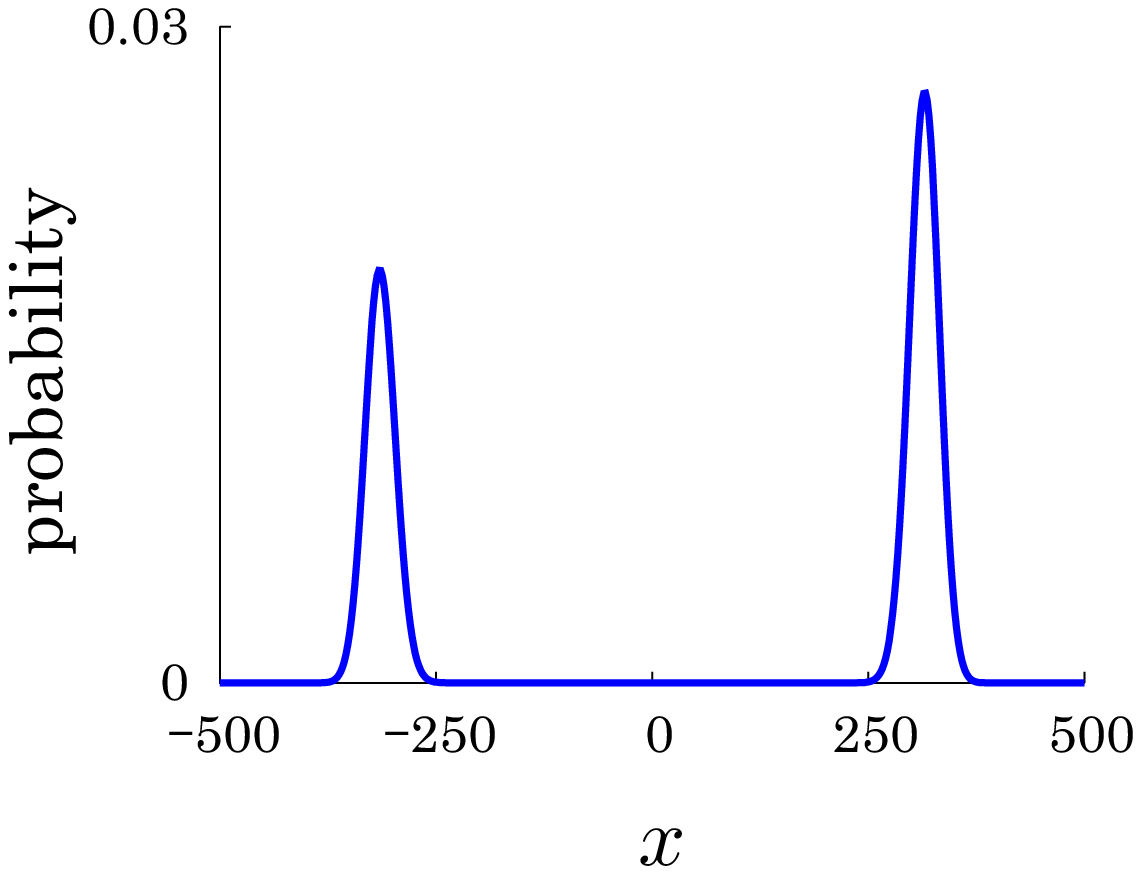}\\[2mm]
  (b) $p=3/4$
  \end{center}
 \end{minipage}
\vspace{5mm}
\caption{(color figure online) $\theta_0=2\pi/7,\,\theta_1=\arcsin(c_0/s_0)$ : Probability distribution of the open quantum walk at time $t=500$ with the initial state $D_0=\ket{0}\otimes (p\ket{0}\bra{0}+(1-p)\ket{1}\bra{1})$.}
\label{fig:1}
\end{center}
\end{figure}

The probability distribution $\mathbb{P}(X_{500}=x)$ holds two peaks in Fig.~\ref{fig:1}.

\begin{figure}[h]
\begin{center}
 \begin{minipage}{50mm}
  \begin{center}
   \includegraphics[scale=0.3]{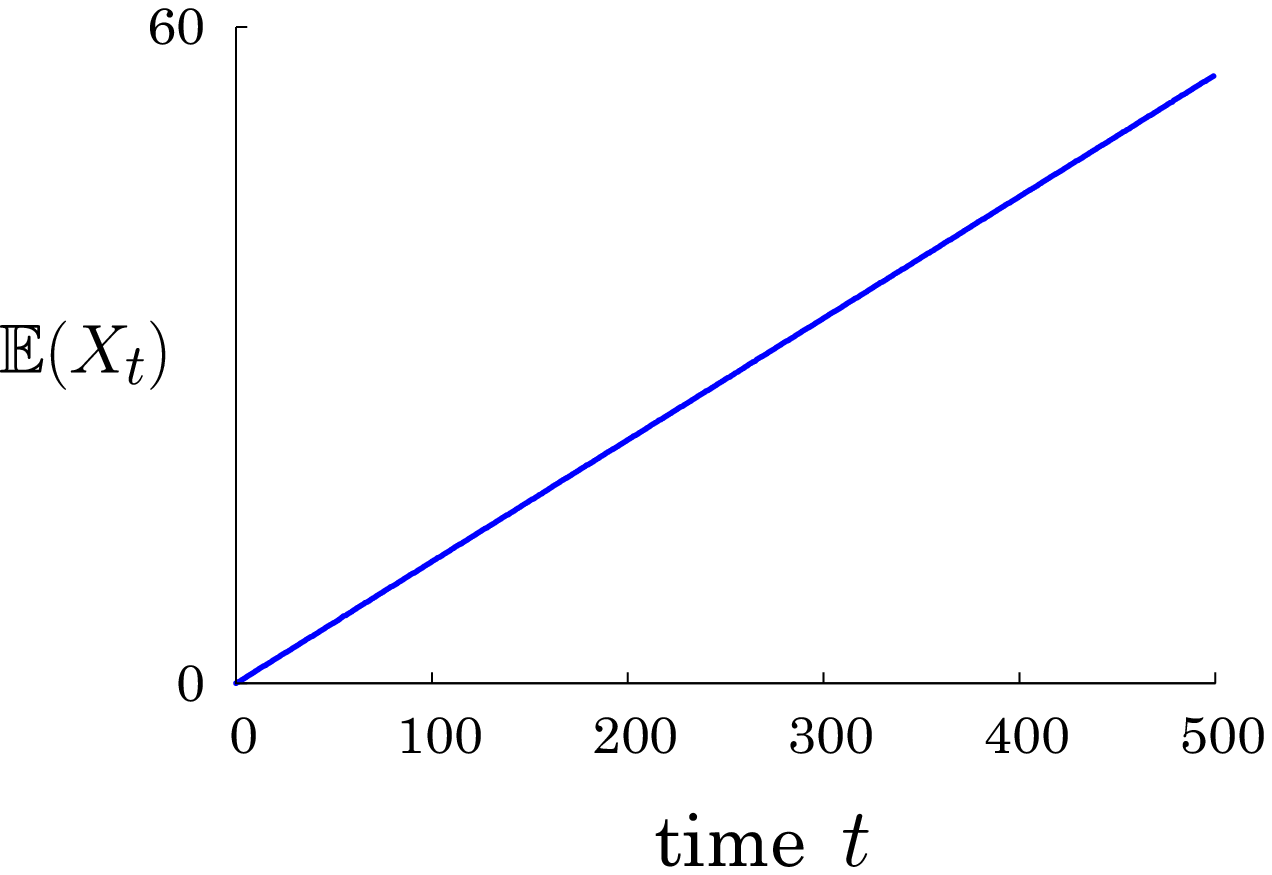}\\[2mm]
  (a) Numerical experiment
  \end{center}
 \end{minipage}
 \begin{minipage}{50mm}
  \begin{center}
   \includegraphics[scale=0.3]{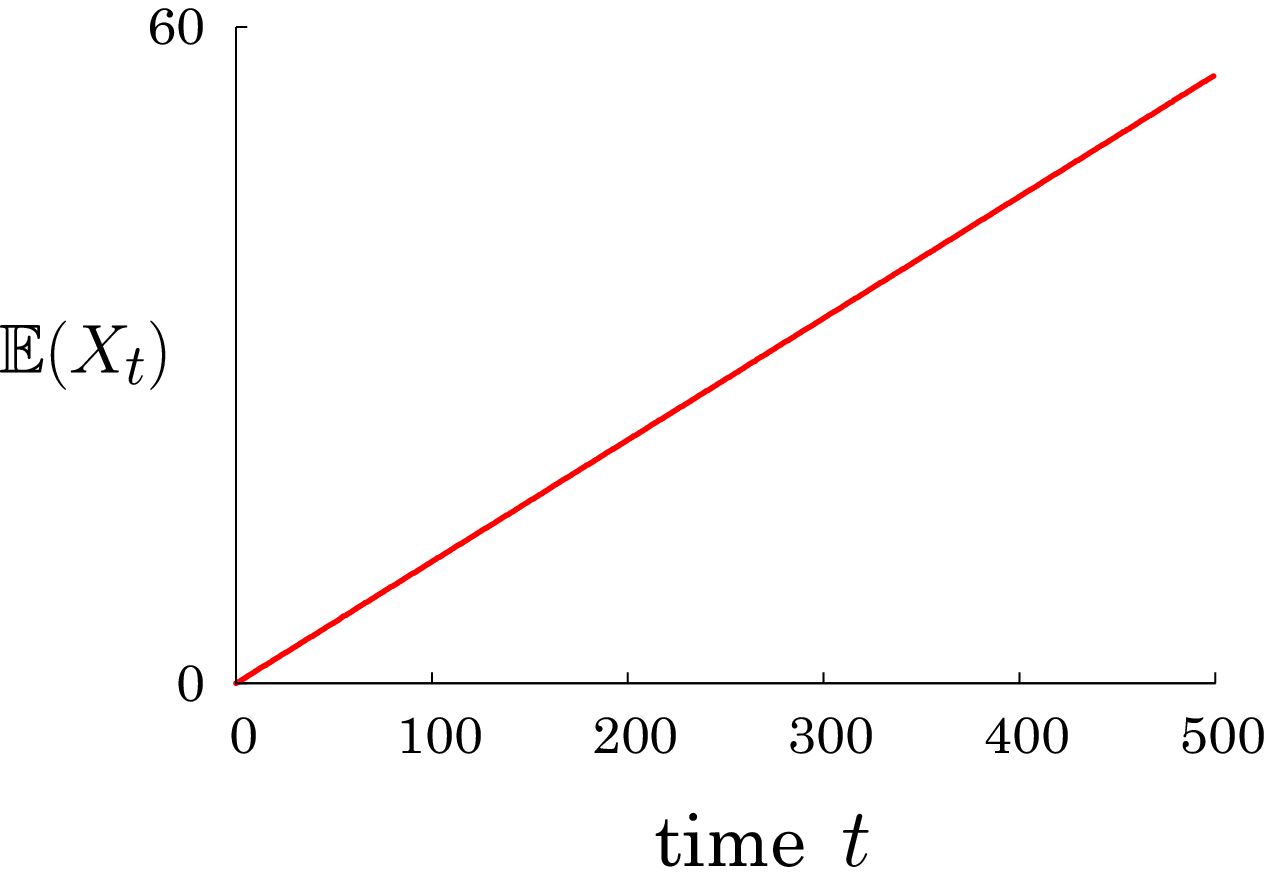}\\[2mm]
  (b) Analytical result
  \end{center}
 \end{minipage}
\caption{(color figure online) $\theta_0=2\pi/7,\,\theta_1=\arcsin(c_0/s_0)$ : The $1$st moment $\mathbb{E}(X_t)$ of the open quantum walk with the initial state $D_0=\ket{0}\otimes (3/4\ket{0}\bra{0}+1/4\ket{1}\bra{1})$.}
\label{fig:2}
\end{center}
\end{figure}

\begin{figure}[h]
\begin{center}
 \begin{minipage}{50mm}
  \begin{center}
   \includegraphics[scale=0.3]{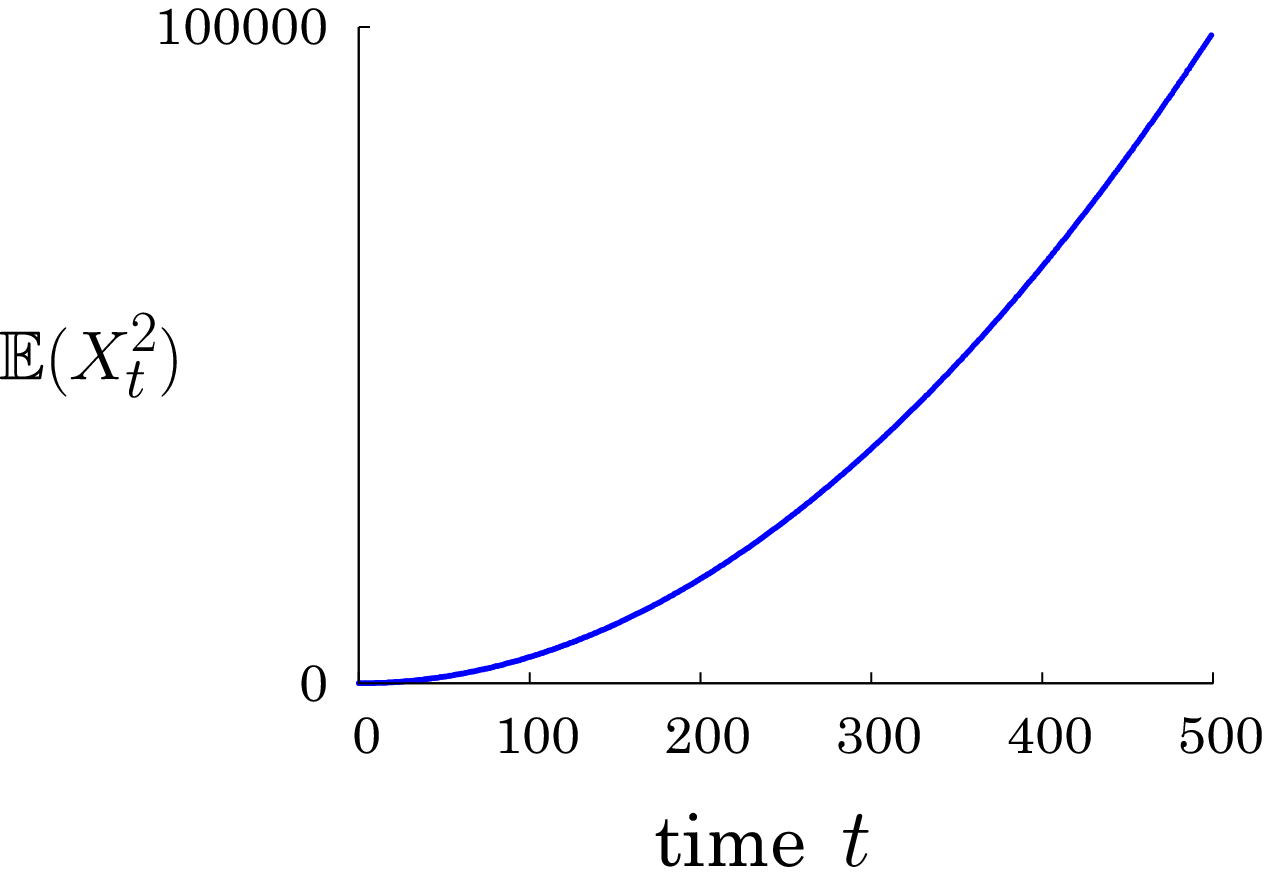}\\[2mm]
  (a) Numerical experiment
  \end{center}
 \end{minipage}
 \begin{minipage}{50mm}
  \begin{center}
   \includegraphics[scale=0.3]{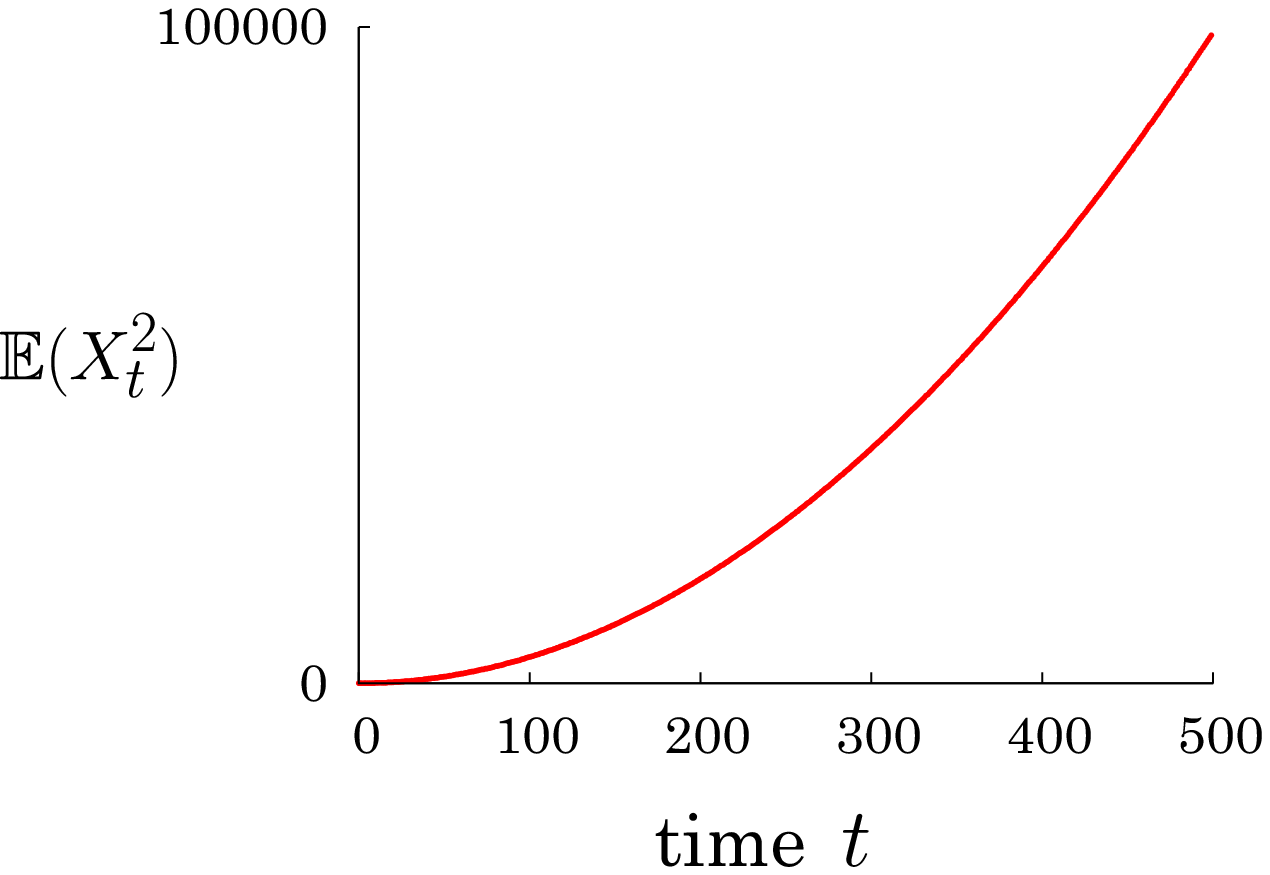}\\[2mm]
  (b) Analytical result
  \end{center}
 \end{minipage}
\caption{(color figure online) $\theta_0=2\pi/7,\,\theta_1=\arcsin(c_0/s_0)$ : The $2$nd moment $\mathbb{E}(X_t^2)$ of the open quantum walk with the initial state $D_0=\ket{0}\otimes (3/4\ket{0}\bra{0}+1/4\ket{1}\bra{1})$.}
\label{fig:3}
\end{center}
\end{figure}

\begin{figure}[h]
\begin{center}
 \begin{minipage}{50mm}
  \begin{center}
   \includegraphics[scale=0.3]{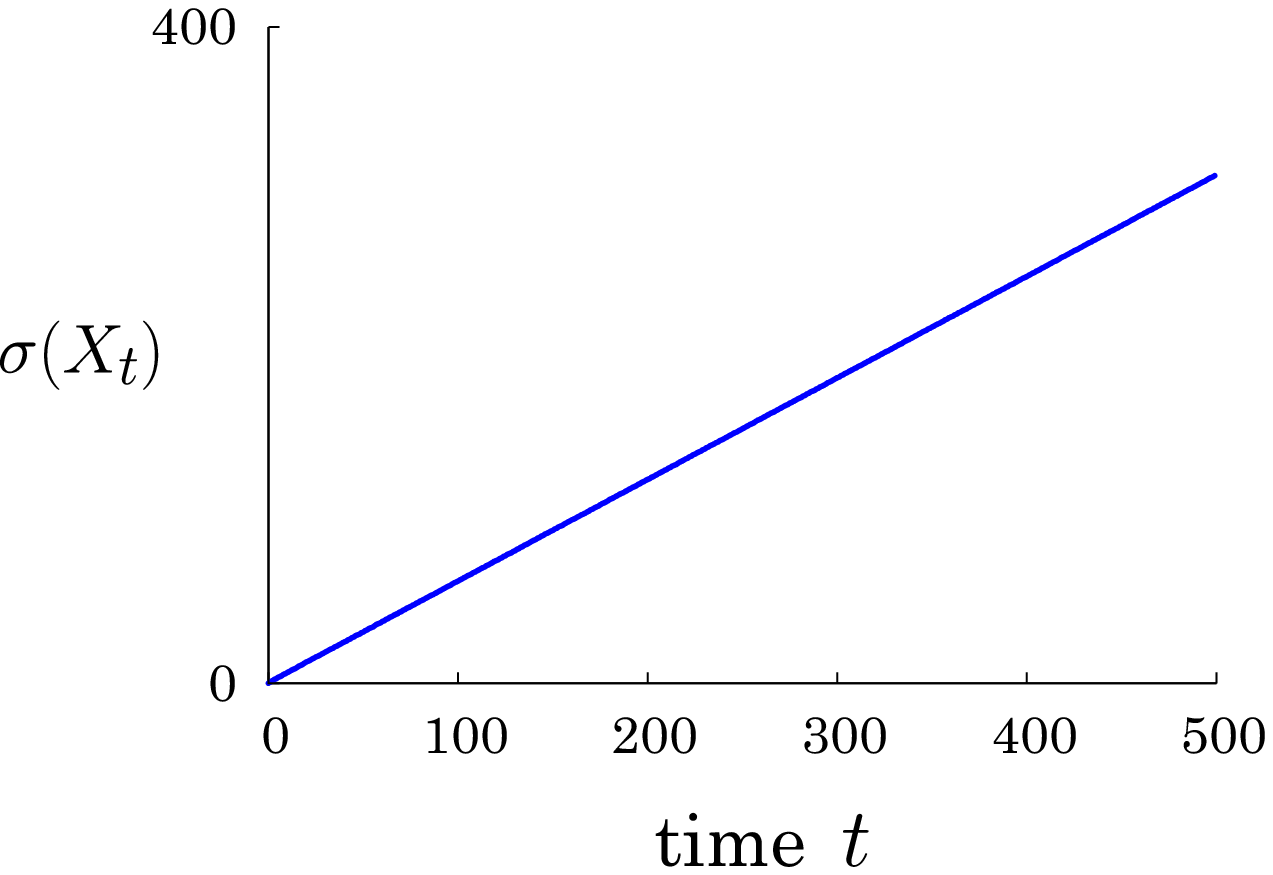}\\[2mm]
  (a) Numerical experiment
  \end{center}
 \end{minipage}
 \begin{minipage}{50mm}
  \begin{center}
   \includegraphics[scale=0.3]{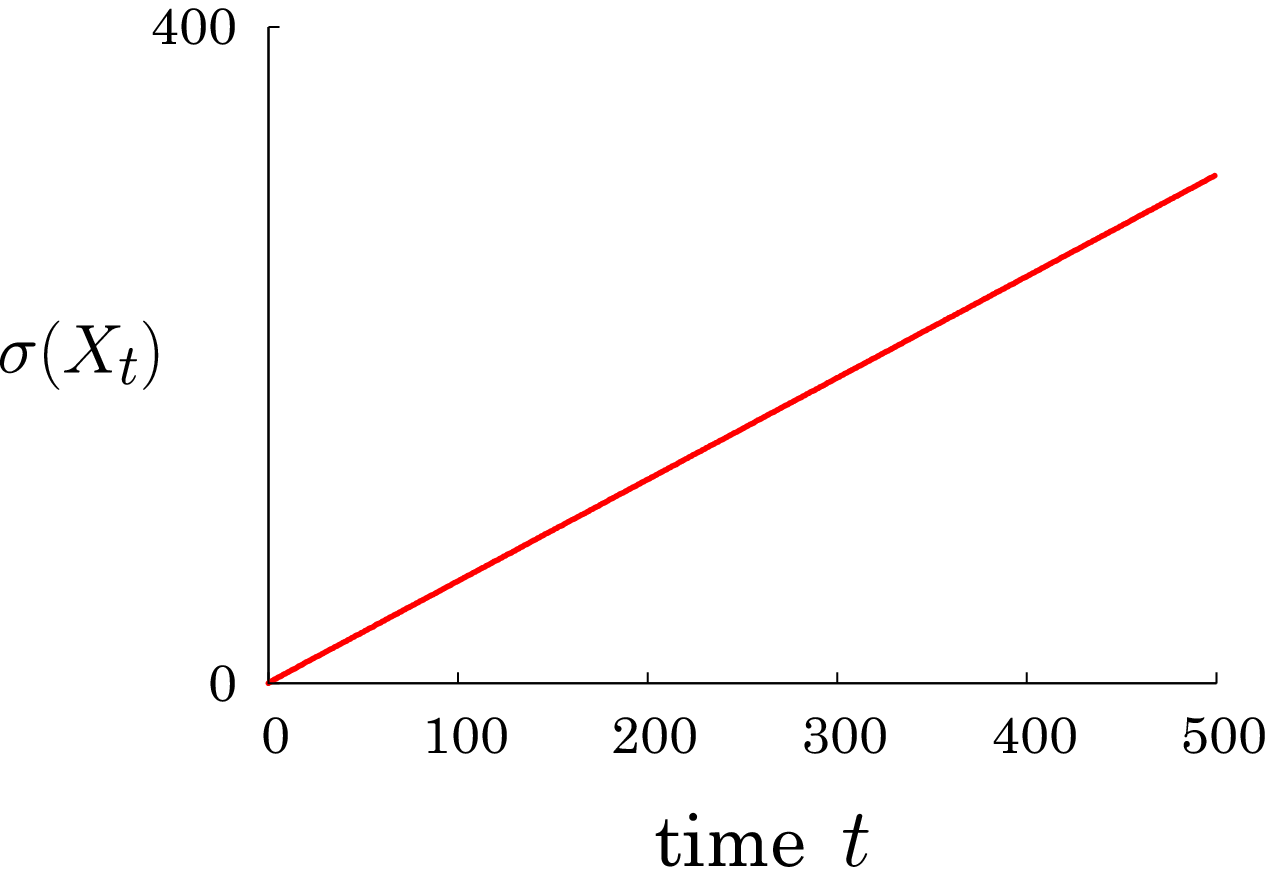}\\[2mm]
  (b) Analytical result
  \end{center}
 \end{minipage}
\caption{(color figure online) $\theta_0=2\pi/7,\,\theta_1=\arcsin(c_0/s_0)$ : The standard deviation $\sigma(X_t)$ of the open quantum walk with the initial state $D_0=\ket{0}\otimes (3/4\ket{0}\bra{0}+1/4\ket{1}\bra{1})$.}
\label{fig:4}
\end{center}
\end{figure}

The 1st moment $\mathbb{E}(X_t)$ linearly increases and the 2nd moment $\mathbb{E}(X_t^2)$ quadratically increases as time $t$ goes up, 
\begin{align}
 \mathbb{E}(X_t)=& (2p-1)s_0^2c_1^2\, t,\label{eq:c0=s0s1-E1}\\[3mm]
 \mathbb{E}(X_t^2)=& s_0^4c_1^2(1+3s_1^2) t^2 + s_0^2\left\{1+s_1^2-s_0^2c_1^2(1+3s_1^2)\right\} t,\label{eq:c0=s0s1-E2}\\[3mm]
 \sigma(X_t)=& \sqrt{s_0^4c_1^2\left\{1+3s_1^2-(2p-1)^2c_1^2\right\} t^2 + s_0^2\left\{1+s_1^2-s_0^2c_1^2(1+3s_1^2)\right\} t},
\end{align}
which are compared to numerical experiments in Figs.~\ref{fig:2}, \ref{fig:3}, and \ref{fig:4}.
\clearpage


\clearpage

\begin{proof}{%
Assuming the parameter $\theta_1\neq 0, \pi$, that is, $s_1\neq 0$, we have the eigenvalues $\lambda_j(k)$ and eigenvectors $\ket{v_j(k)}$,    
\begin{align}
 \lambda_1(k)=& s_0^2\left\{(1+s_1^2)\cos k - i\, c_1\sqrt{1+3s_1^2}\,\sin k\right\},\\[3mm]
 \ket{v_1(k)}=&
 \begin{bmatrix}
  \left\{2s_1^2(1+e^{2ik})+c_1\left(c_1+\sqrt{1+3s_1^2}\,\right)\right\}\left(c_1+\sqrt{1+3s_1^2}\,\right)\\[3mm]
  4s_1^2 e^{ik}\left(\sqrt{1+3s_1^2}\,\cos k - i\, c_1\sin k\right)\\[3mm]
  2s_1\left\{2s_1^2(1+e^{2ik})+c_1\left(c_1+\sqrt{1+3s_1^2}\,\right)\right\}\\[3mm]
  2s_1\left\{2s_1^2(1+e^{2ik})+c_1\left(c_1+\sqrt{1+3s_1^2}\,\right)\right\}
 \end{bmatrix},
\end{align}
\begin{align}
 \lambda_2(k)=& s_0^2\left\{(1+s_1^2)\cos k + i\, c_1\sqrt{1+3s_1^2}\,\sin k\right\},\\[3mm]
 \ket{v_2(k)}=&
 \begin{bmatrix}
  \left\{2s_1^2(1+e^{2ik})+c_1\left(c_1-\sqrt{1+3s_1^2}\,\right)\right\}\left(c_1-\sqrt{1+3s_1^2}\,\right)\\[3mm]
  -4s_1^2 e^{ik}\left(\sqrt{1+3s_1^2}\,\cos k + i\, c_1\sin k\right)\\[3mm]
  2s_1\left\{2s_1^2(1+e^{2ik})+c_1\left(c_1-\sqrt{1+3s_1^2}\,\right)\right\}\\[3mm]
  2s_1\left\{2s_1^2(1+e^{2ik})+c_1\left(c_1-\sqrt{1+3s_1^2}\,\right)\right\}
 \end{bmatrix},
\end{align}
\begin{align}
 & \lambda_3(k)= \lambda_4(k)= -2s_0^2s_1^2\cos k,\\[3mm]
 & \ket{v_3(k)}=
 \begin{bmatrix}
  0\\0\\-1\\1
 \end{bmatrix},\quad
 \ket{v_4(k)}=
 \begin{bmatrix}
  -2s_1\\2s_1\\c_1\\c_1
 \end{bmatrix}.
\end{align}
Note that
\begin{equation}
 \lambda_1(0)=\lambda_2(0)
 = s_0^2(1+s_1^2)
 = s_0^2+s_0^2s_1^2
 = s_0^2+c_0^2=1,
\end{equation}
and the eigenvectors $\ket{v_3(k)}$ and $\ket{v_4(k)}$ are orthogonal to $\ket{00}+\ket{01}$, which means $w_3(k)=w_4(k)=0$.
Since the eigenvectors are orthogonal to each other, we find
\begin{equation}
 a_j(k)=\frac{\braket{v_j(k)|\hat{\psi}_0(k)}}{\braket{v_j(k)|v_j(k)}}.
\end{equation}
The 1st and 2nd moments, therefore, arrive in the same forms as Eqs.~\eqref{eq:c0=s0s1-E1} and \eqref{eq:c0=s0s1-E2}.

\bigskip

On the other hand, the operation $\hat{U}(k)$ becomes a diagonal matrix
\begin{equation}
 \hat{U}(k)
  =\begin{bmatrix}
    e^{-ik} & 0 & 0 & 0\\
    0 & e^{ik} & 0 & 0\\
    0 & 0 & 0 & 0\\
    0 & 0 & 0 & 0
   \end{bmatrix},
\end{equation}
if the value of parameter $\theta_1$ is fixed at $0$ or $\pi$.
Then, the system of the open quantum walk at time $t$ results in
\begin{equation}
 \ket{\hat\psi_t(k)}=e^{-ikt} p\ket{00} + e^{ikt} (1-p)\ket{01},
\end{equation}
following
\begin{equation}
 \mathbb{P}(X_t=t)=p,\quad \mathbb{P}(X_t=-t)=1-p.
\end{equation}
The 1st and 2nd moments
\begin{equation}
 \mathbb{E}(X_t)=(2p-1)\,t,\quad \mathbb{E}(X_t^2)=t^2,
\end{equation}
are allowed to be included in the representations of Eqs.~\eqref{eq:c0=s0s1-E1} and \eqref{eq:c0=s0s1-E2}. 
}\end{proof}

\clearpage
\subsection{Case: $c_0\neq s_0s_1$}
\label{subsec:c0!=s0s1}

\begin{enumerate}
 \item Case: $s_0c_1=0$
       
       The probability distribution $\mathbb{P}(X_{500}=x)$ seems to be like a Gaussian distribution in numerical experiments, as Fig.~\ref{fig:5} shows.
\begin{figure}[h]
\begin{center}
 \begin{minipage}{50mm}
  \begin{center}
   \includegraphics[scale=0.4]{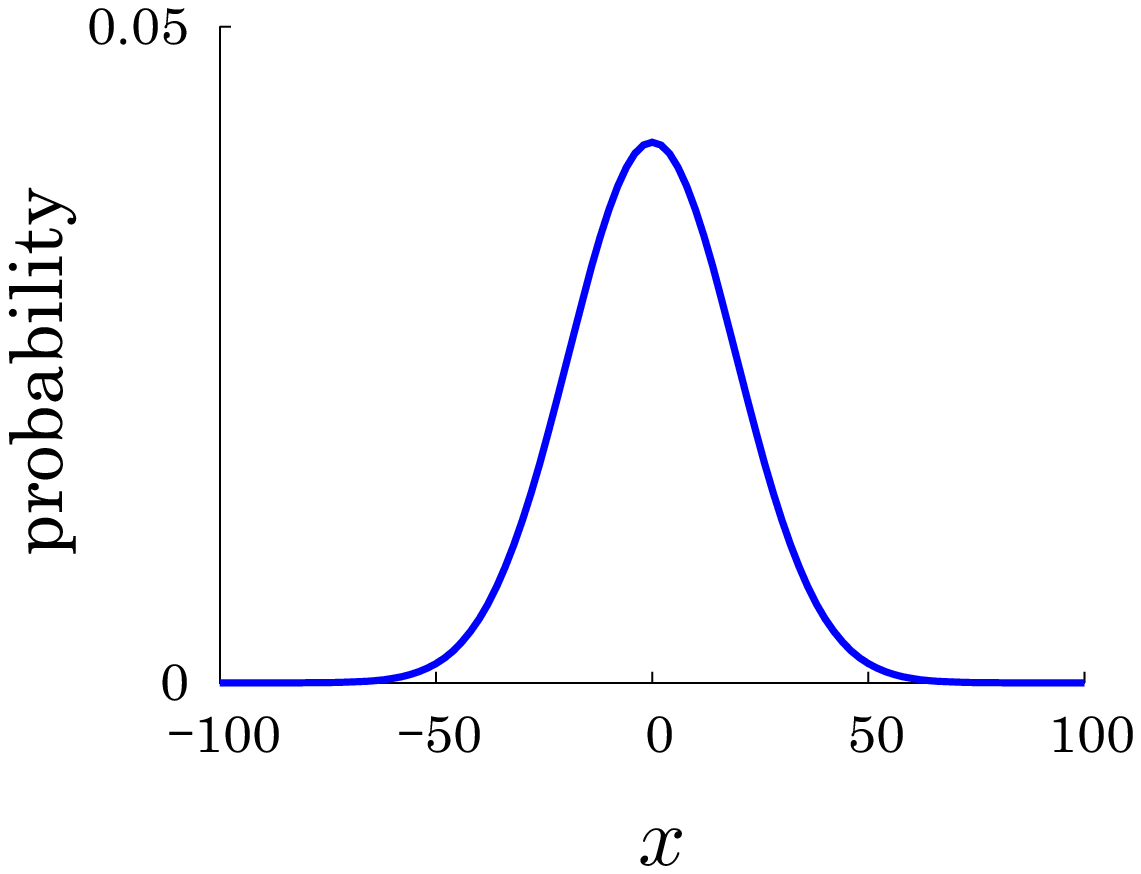}\\[2mm]
  (a) $p=1/2$
  \end{center}
 \end{minipage}
 \begin{minipage}{50mm}
  \begin{center}
   \includegraphics[scale=0.4]{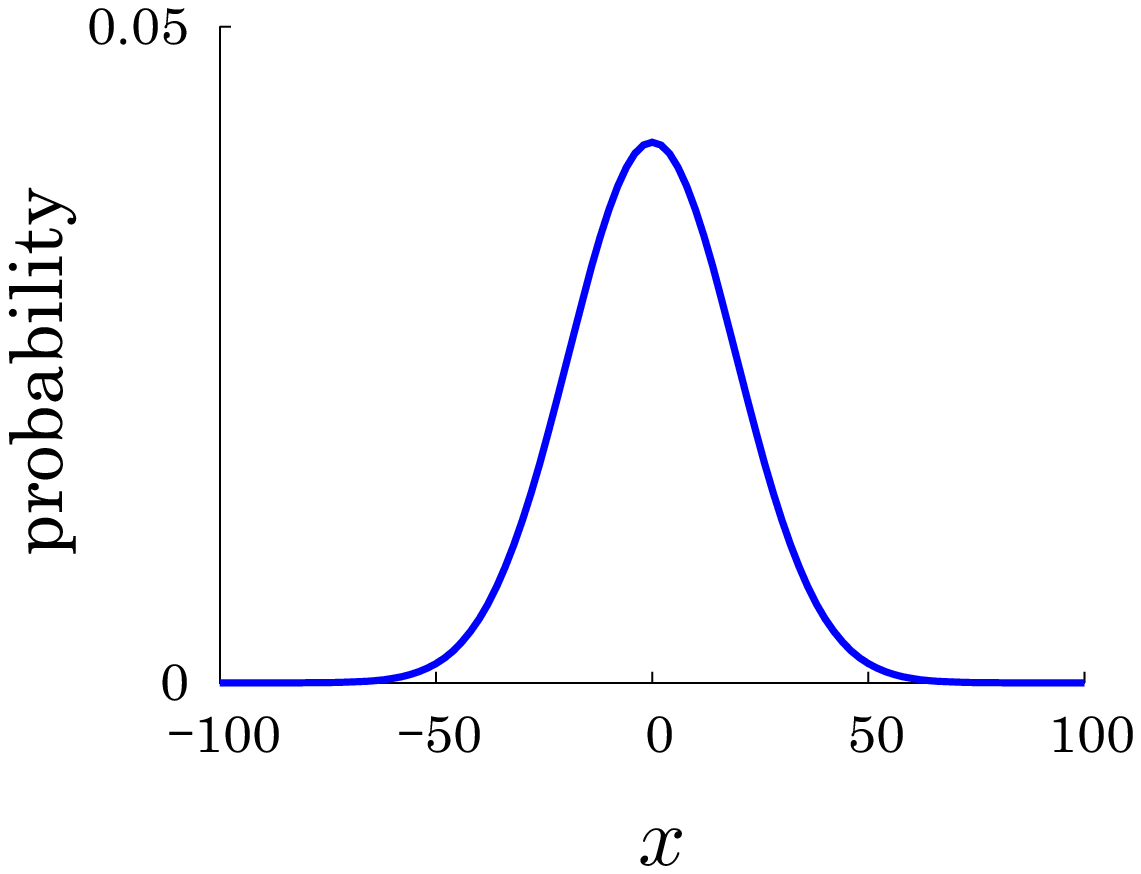}\\[2mm]
  (b) $p=3/4$
  \end{center}
 \end{minipage}
\vspace{5mm}
\caption{(color figure online) $\theta_0=\pi/3,\,\theta_1=\pi/2$ : Probability distribution of the open quantum walk at time $t=500$ with the initial state $D_0=\ket{0}\otimes (p\ket{0}\bra{0}+(1-p)\ket{1}\bra{1})$.}
\label{fig:5}
\end{center}
\end{figure}

       The 1st and 2nd moments have two representations according to the value of time $t$,
       \begin{align}
	\mathbb{E}(X_t)=& \left\{\begin{array}{ll}
			   0 & (t=0,2,4,\ldots)\\[1mm]
			    (2p-1)(2c_0^2-1) & (t=1,3,5,\ldots)
				 \end{array}\right.,\label{eq:c0!=s0s1-s0c1=0-E1}\\[3mm]
	\mathbb{E}(X_t^2)=& \left\{\begin{array}{ll}
			     4c_0^2s_0^2\, t & (t=0,2,4,\ldots)\\[1mm]
			      4c_0^2s_0^2\,t + (2c_0^2-1)^2 & (t=1,3,5,\ldots)
				   \end{array}\right.,\label{eq:c0!=s0s1-s0c1=0-E2}\\[3mm]
	\sigma(X_t)=& \left\{\begin{array}{ll}
		       2|c_0s_0|\sqrt{t} & (t=0,2,4,\ldots)\\[1mm]
			2\sqrt{c_0^2s_0^2\,t + p(1-p)(2c_0^2-1)^2} & (t=1,3,5,\ldots)
			     \end{array}\right.,
       \end{align}
       which match numerical experiments in Figs.~\ref{fig:6}, \ref{fig:7}, and \ref{fig:8}.

\clearpage

\begin{figure}[h]
\begin{center}
 \begin{minipage}{50mm}
  \begin{center}
   \includegraphics[scale=0.3]{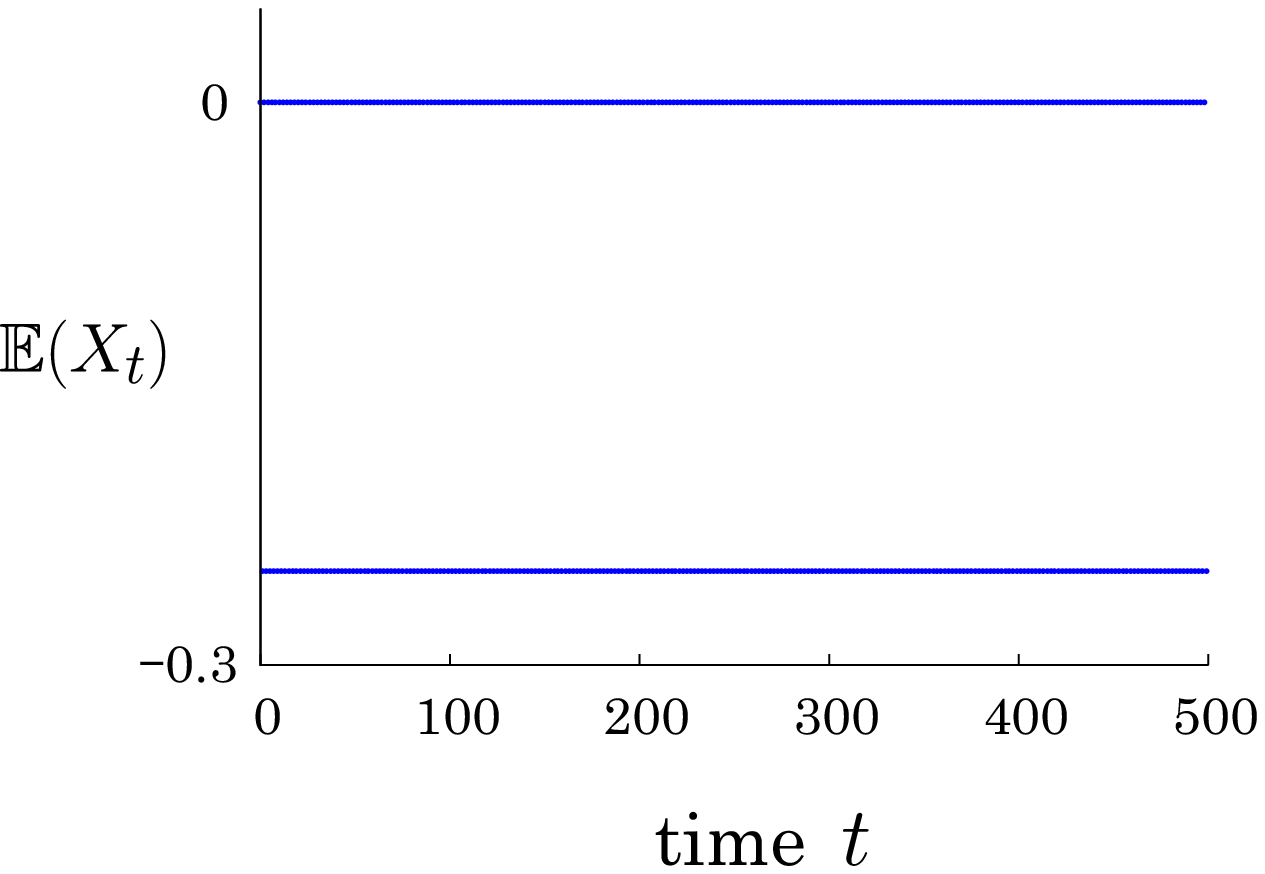}\\[2mm]
  (a) Numerical experiment
  \end{center}
 \end{minipage}
 \begin{minipage}{50mm}
  \begin{center}
   \includegraphics[scale=0.3]{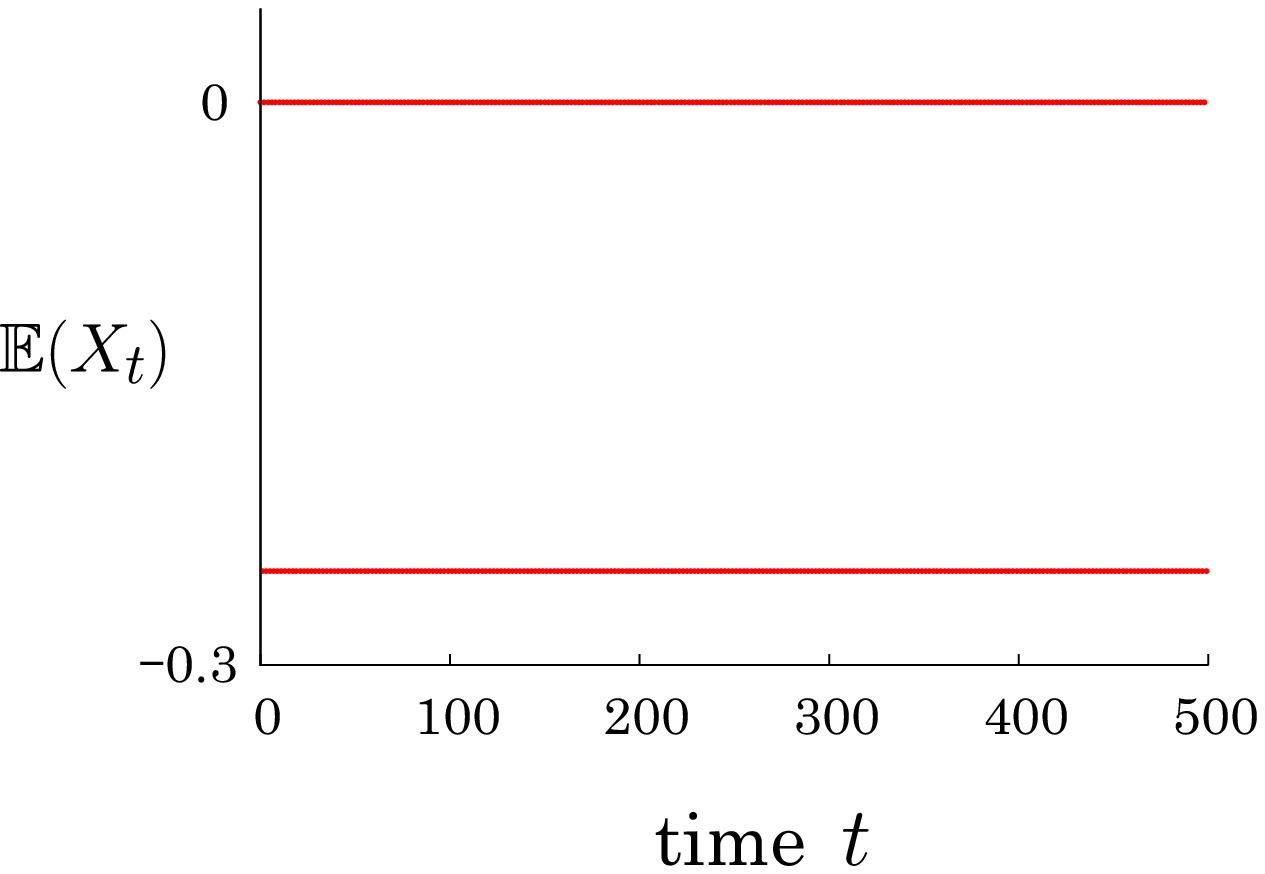}\\[2mm]
  (b) Analytical result
  \end{center}
 \end{minipage}
\caption{(color figure online) $\theta_0=\pi/3,\,\theta_1=\pi/2$ : The $1$st moment $\mathbb{E}(X_t)$ of the open quantum walk with the initial state $D_0=\ket{0}\otimes (3/4\ket{0}\bra{0}+1/4\ket{1}\bra{1})$.}
\label{fig:6}
\end{center}
\end{figure}
\begin{figure}[h]
\begin{center}
 \begin{minipage}{50mm}
  \begin{center}
   \includegraphics[scale=0.3]{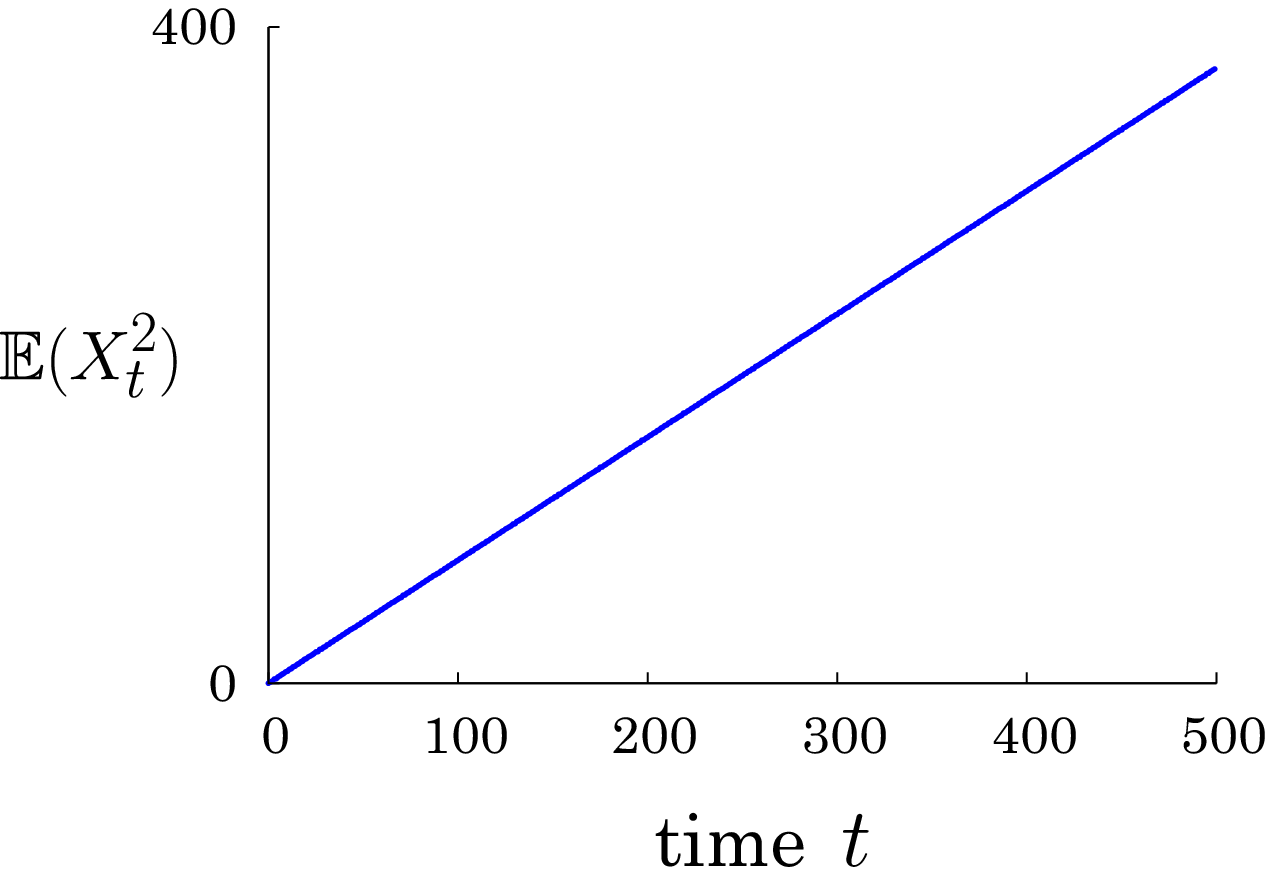}\\[2mm]
  (a) Numerical experiment
  \end{center}
 \end{minipage}
 \begin{minipage}{50mm}
  \begin{center}
   \includegraphics[scale=0.3]{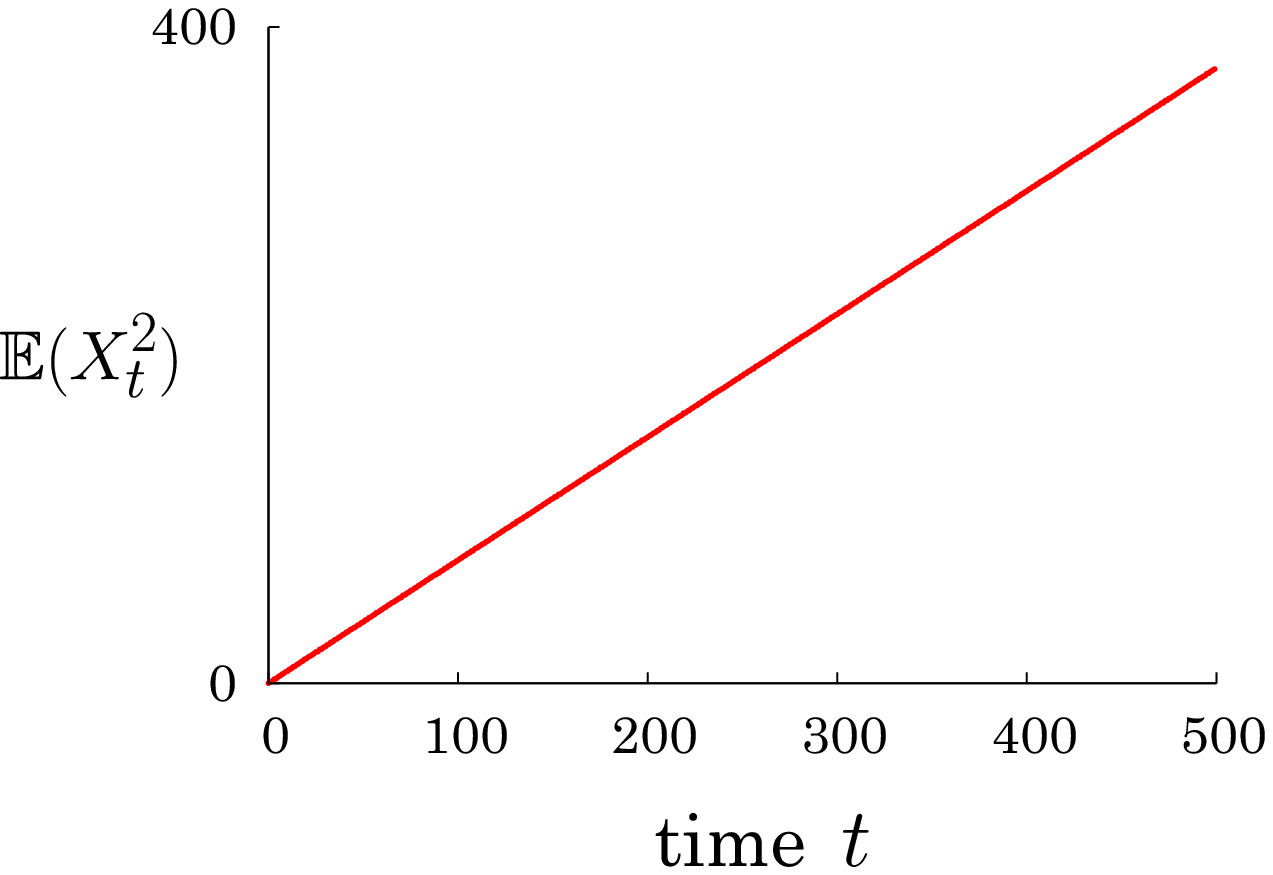}\\[2mm]
  (b) Analytical result
  \end{center}
 \end{minipage}
\caption{(color figure online) $\theta_0=\pi/3,\,\theta_1=\pi/2$ : The $2$nd moment $\mathbb{E}(X_t^2)$ of the open quantum walk with the initial state $D_0=\ket{0}\otimes (3/4\ket{0}\bra{0}+1/4\ket{1}\bra{1})$.}
\label{fig:7}
\end{center}
\end{figure}
\begin{figure}[h]
\begin{center}
 \begin{minipage}{50mm}
  \begin{center}
   \includegraphics[scale=0.3]{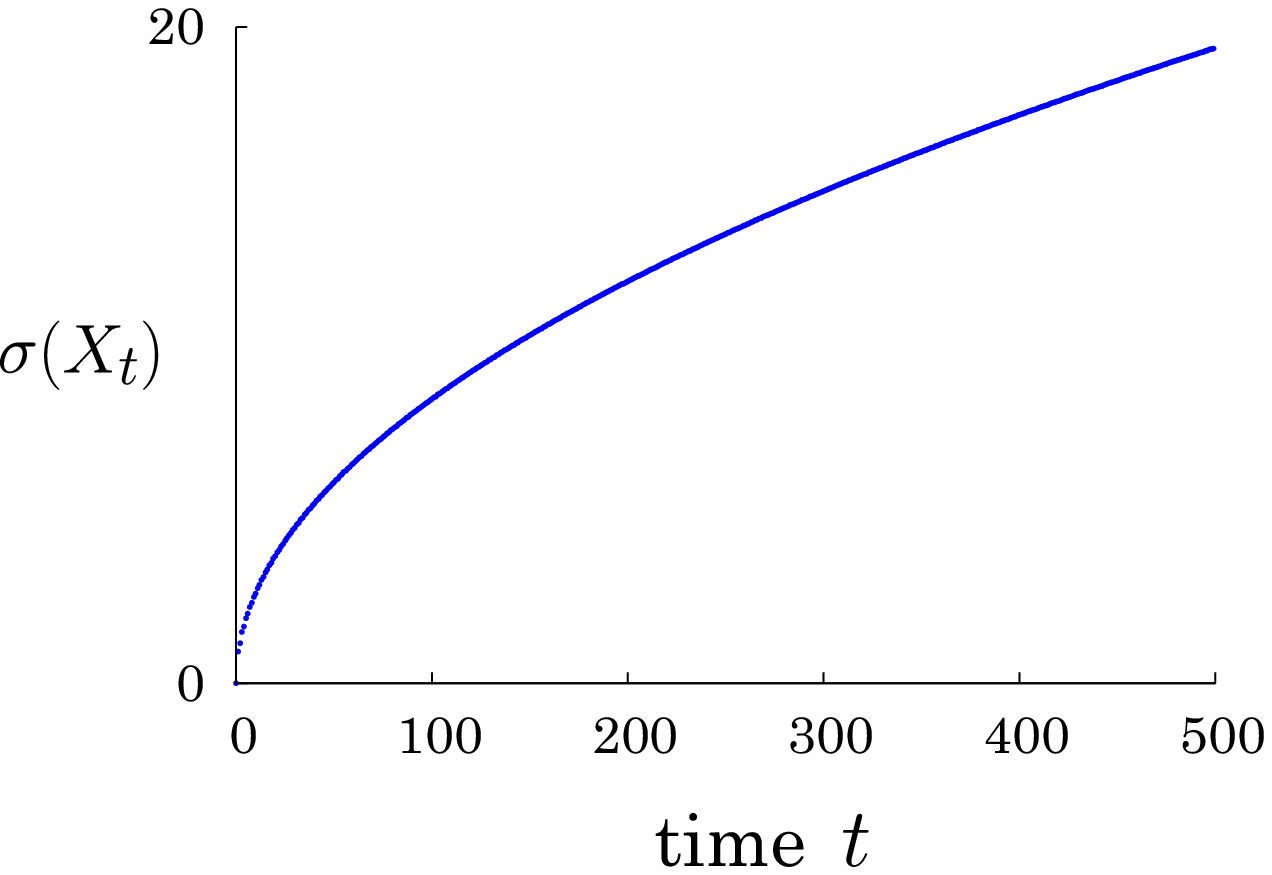}\\[2mm]
  (a) Numerical experiment
  \end{center}
 \end{minipage}
 \begin{minipage}{50mm}
  \begin{center}
   \includegraphics[scale=0.3]{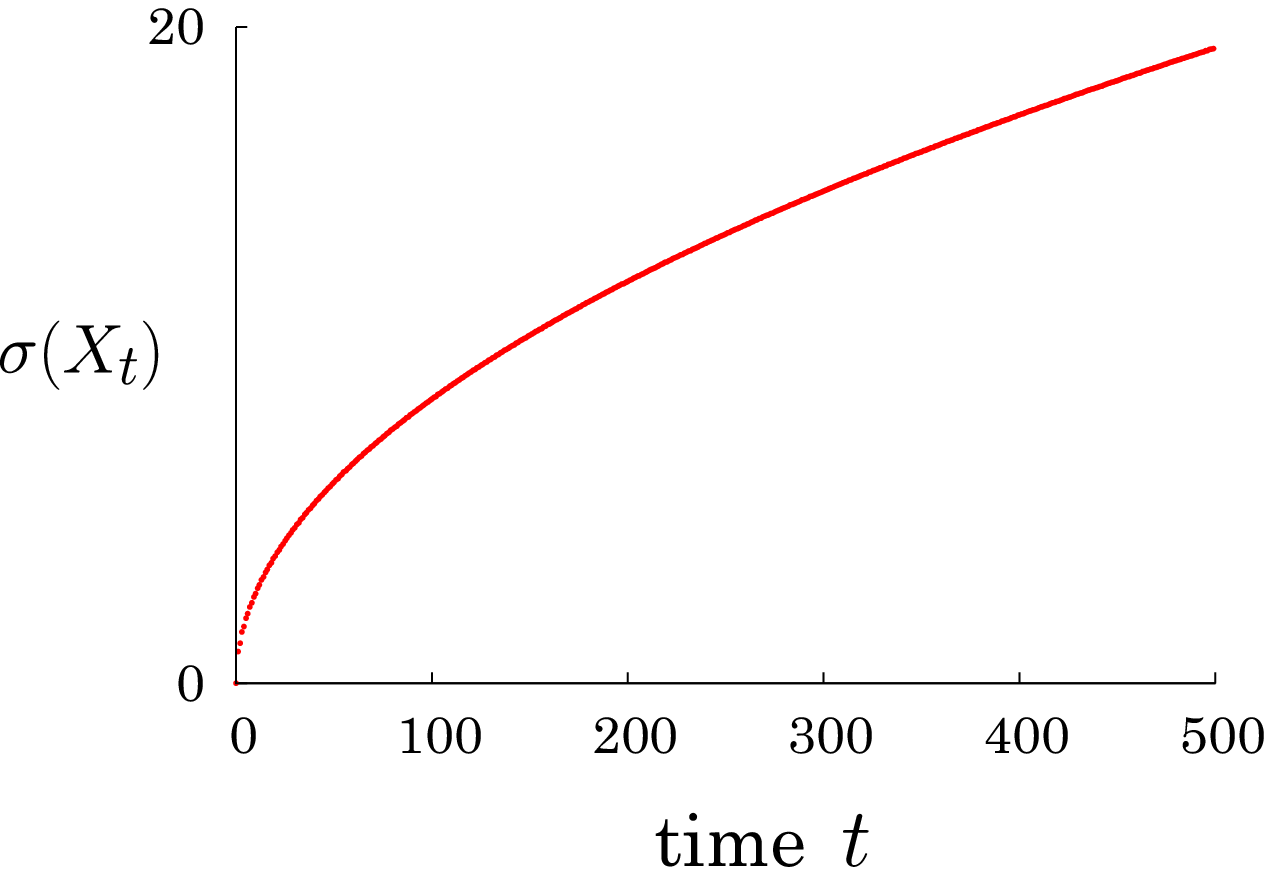}\\[2mm]
  (b) Analytical result
  \end{center}
 \end{minipage}
\caption{(color figure online) $\theta_0=\pi/3,\,\theta_1=\pi/2$ : The standard deviation $\sigma(X_t)$ of the open quantum walk with the initial state $D_0=\ket{0}\otimes (3/4\ket{0}\bra{0}+1/4\ket{1}\bra{1})$.}
\label{fig:8}
\end{center}
\end{figure}

\clearpage

\begin{proof}{%
The operation $\hat{U}(k)$ becomes
\begin{align}
 \hat{U}(k)
  =& (e^{ik}c_0^2+e^{-ik}s_0^2s_1^2)\ket{00}\bra{01}\nonumber\\
  & +(e^{-ik}c_0^2+e^{ik}s_0^2s_1^2)\ket{01}\bra{00}\nonumber\\
  & +(e^{-ik}c_0s_0s_1+e^{ik}c_0s_0s_1)\ket{10}\bra{11}\nonumber\\
  & +(e^{-ik}c_0s_0s_1+e^{ik}c_0s_0s_1)\ket{11}\bra{10},
\end{align}
and the system of the open quantum walk at time $t$ results in
\begin{align}
 \ket{\hat\psi_t(k)}
 =&
  \begin{bmatrix}
   \begin{bmatrix}
    0 & e^{ik}c_0^2+e^{-ik}s_0^2s_1^2\\
    e^{-ik}c_0^2+e^{ik}s_0^2s_1^2 & 0
   \end{bmatrix}^{t}
   \begin{bmatrix}
    p\\1-p
   \end{bmatrix}\\
   0\\0
  \end{bmatrix}\nonumber\\
 =&
  \begin{bmatrix}
   \hat{V}(k)^t\,\ket{\phi}\\
   0\\0
  \end{bmatrix},\label{eq:201221-1}
\end{align}
where $\ket{\phi}=p\ket{0}+(1-p)\ket{1}$ and
\begin{equation}
 \hat{V}(k)=
 \begin{bmatrix}
  0 & e^{ik}c_0^2+e^{-ik}s_0^2s_1^2\\
  e^{-ik}c_0^2+e^{ik}s_0^2s_1^2 & 0
 \end{bmatrix}.
\end{equation}
Equation~\eqref{eq:201221-1} allows us to analyze the vector $\ket{\hat\psi_t(k)}\in\mathbb{C}^4$ in the reduced space $\mathbb{C}^2$.
The $2\times 2$ matrix $\hat{V}(k)$ holds two eigenvalues
\begin{equation}
 \nu_j(k)=-(-1)^j\sqrt{1-4c_0^2s_0^2\sin^2k}\quad (j=1,2).
\end{equation}
Note that $\nu_1(0)=1$ and $\nu_2(0)=-1$.
A possible eigenvector associated to the eigenvalue $\nu_j(k)$, represented by $\ket{\xi_j(k)}$, is of the form
\begin{equation}
 \ket{\xi_j(k)}=
  \begin{bmatrix}
   -(-1)^j \sqrt{1-4c_0^2s_0^2\sin^2 k}\,\,\\[3mm]
   e^{-ik}c_0^2+e^{ik}s_0^2
  \end{bmatrix}.
\end{equation}

Decomposing the reduced initial state $\ket{\phi}=p\ket{0}+(1-p)\ket{1}$ in the eigenspace,
\begin{equation}
 \ket{\phi}=b_1(k)\ket{\xi_1(k)}+b_2(k)\ket{\xi_2(k)},
\end{equation}
we have
\begin{align}
 \mathbb{E}(X_t)
 = & \Bigl(\bra{00}+\bra{01}\Bigr)\left(i\frac{d}{dk} \ket{\hat\psi_t(k)}\right)\bigg|_{k=0}\nonumber\\
 = & \sum_{j=1}^2 i\,t\, \nu_j(0)^{t-1}\nu'_j(0)b_j(0)z_j(0) + i\,\nu_j(0)^t\Bigl(b_j(k)z_j(k)\Bigr)'\Big|_{k=0},\label{eq:201221-2}\\
 \mathbb{E}(X_t^2)
 = & \Bigl(\bra{00}+\bra{01}\Bigr)\left(i^2\frac{d^2}{dk^2} \ket{\hat\psi_t(k)}\right)\bigg|_{k=0}\nonumber\\
 = & \sum_{j=1}^2 -\, t^2\, \nu_j(0)^{t-2}\left(\nu'_j(0)\right)^2 b_j(0)z_j(0)\nonumber\\
 & +\,t\,\biggl[\nu_j(0)^{t-2}\left(\nu'_j(0)\right)^2 b_j(0)z_j(0)\nonumber\\
 & -\,2\nu_j(0)^{t-1}\nu'_j(0)\Bigl(b_j(k)z_j(k)\Bigr)'\Big|_{k=0}\nonumber\\
 & -\,\nu_j(0)^{t-1}\nu''_j(0)b_j(0)z_j(0)\biggr]\nonumber\\
 & -\,\nu_j(0)^t\Bigl(b_j(k)z_j(k)\Bigr)''\Big|_{k=0},\label{eq:201221-3}
\end{align}
where
\begin{equation}
 z_j(k)=\Bigl(\bra{0}+\bra{1}\Bigr)\ket{\xi_j(k)}\quad (j=1,2).
\end{equation}
Since the eigenvectors $\ket{\xi_1(k)}$ and $\ket{\xi_2(k)}$ are orthogonal to each other, we have
\begin{equation}
 b_j(k)=\frac{\braket{\xi_j(k)|\phi}}{\braket{\xi_j(k)|\xi_j(k)}},
\end{equation}
and get the 1st and 2nd moments which were shown in Eqs.~\eqref{eq:c0!=s0s1-s0c1=0-E1} and \eqref{eq:c0!=s0s1-s0c1=0-E2}.
}\end{proof}

\clearpage

 \item Case: $s_0c_1\neq 0$

       \begin{enumerate}
	\item Case: $c_0=0$

	      Figure~\ref{fig:9} depicts two examples of the probability distribution $\mathbb{P}(X_{500}=x)$ if the parameters $\theta_0$ and $\theta_1$ are set at values such that $s_0c_1\neq 0$ and $c_0=0$.
\begin{figure}[h]
\begin{center}
 \begin{minipage}{50mm}
  \begin{center}
   \includegraphics[scale=0.4]{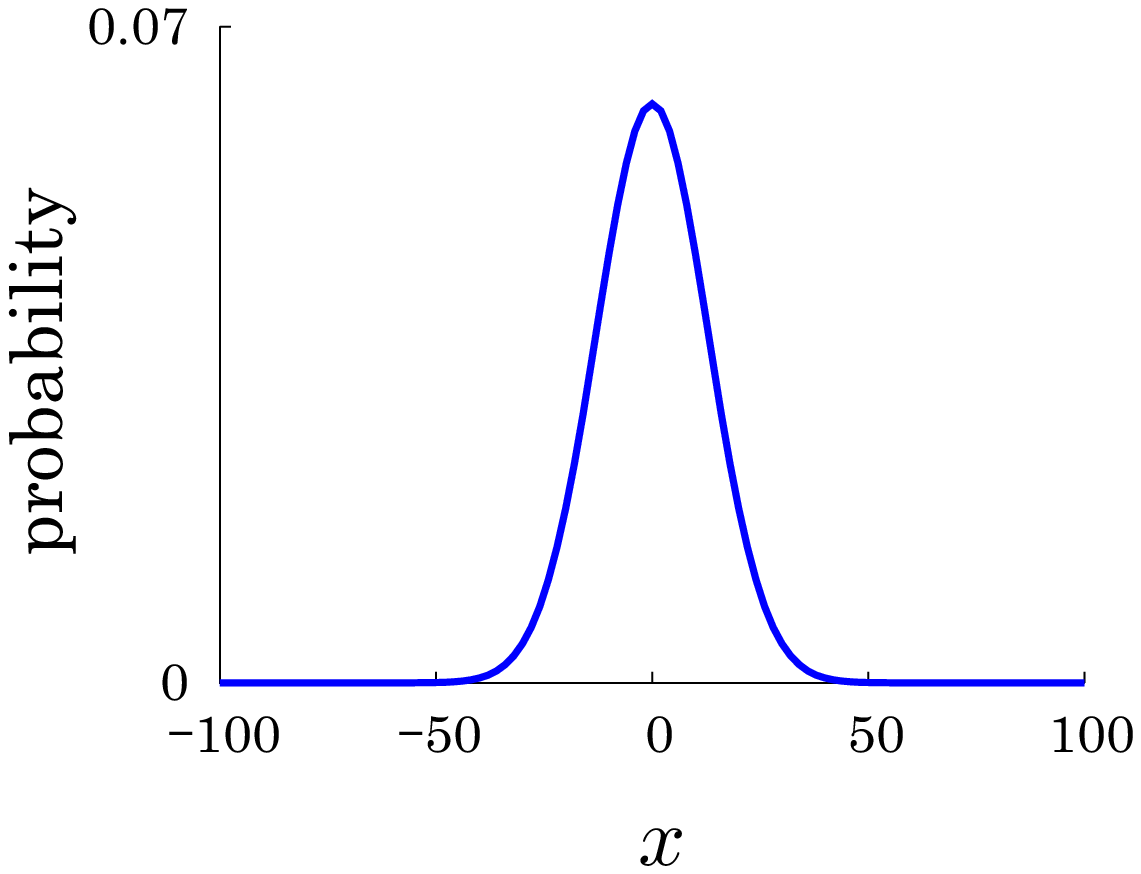}\\[2mm]
  (a) $p=1/2$
  \end{center}
 \end{minipage}
 \begin{minipage}{50mm}
  \begin{center}
   \includegraphics[scale=0.4]{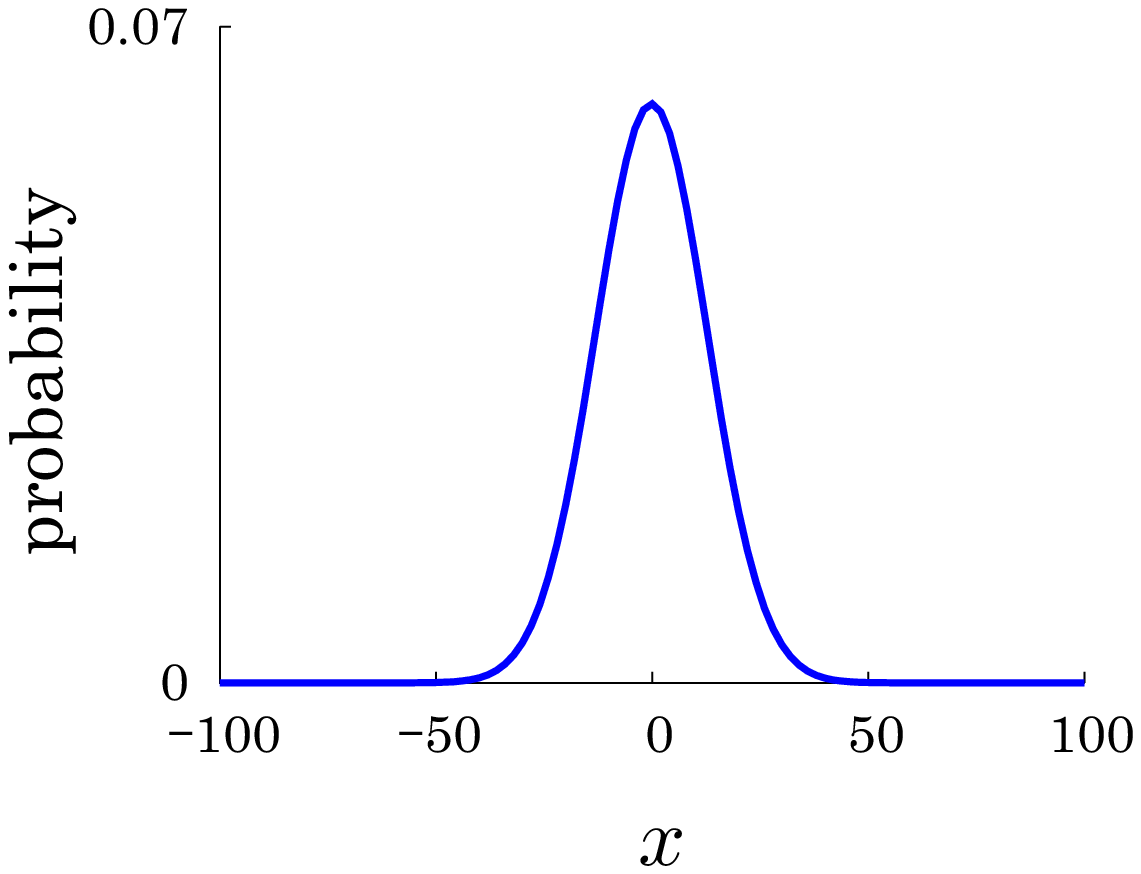}\\[2mm]
  (b) $p=3/4$
  \end{center}
 \end{minipage}
\vspace{5mm}
\caption{(color figure online) $\theta_0=\pi/2,\,\theta_1=\pi/3$ : Probability distribution of the open quantum walk at time $t=500$ with the initial state $D_0=\ket{0}\otimes (p\ket{0}\bra{0}+(1-p)\ket{1}\bra{1})$.}
\label{fig:9}
\end{center}
\end{figure}

	      We have approximations for large values of time $t$,
	      \begin{align}
	       \mathbb{E}(X_t)\sim & \frac{(2p-1)(1-2s_1^2)}{2s_1^2},\\[3mm]
	       \mathbb{E}(X_t^2)\sim & \frac{c_1^2}{s_1^2}\,t - \frac{1-2s_1^2}{2s_1^4}\, \sim \frac{c_1^2}{s_1^2}\,t,\\[3mm]
	       \sigma(X_t)\sim & \sqrt{\frac{c_1^2}{s_1^2}\,t - \frac{(1-2s_1^2)\left\{2+(2p-1)^2(1-2s_1^2)\right\}}{4s_1^4}}\, \sim \frac{|c_1|}{|s_1|}\sqrt{t}.
	      \end{align}
	      These approximations can be checked in Figs.~\ref{fig:10}, \ref{fig:11}, and \ref{fig:12}, compared to numerical experiments.

\clearpage

\begin{figure}[h]
\begin{center}
 \begin{minipage}{50mm}
  \begin{center}
   \includegraphics[scale=0.3]{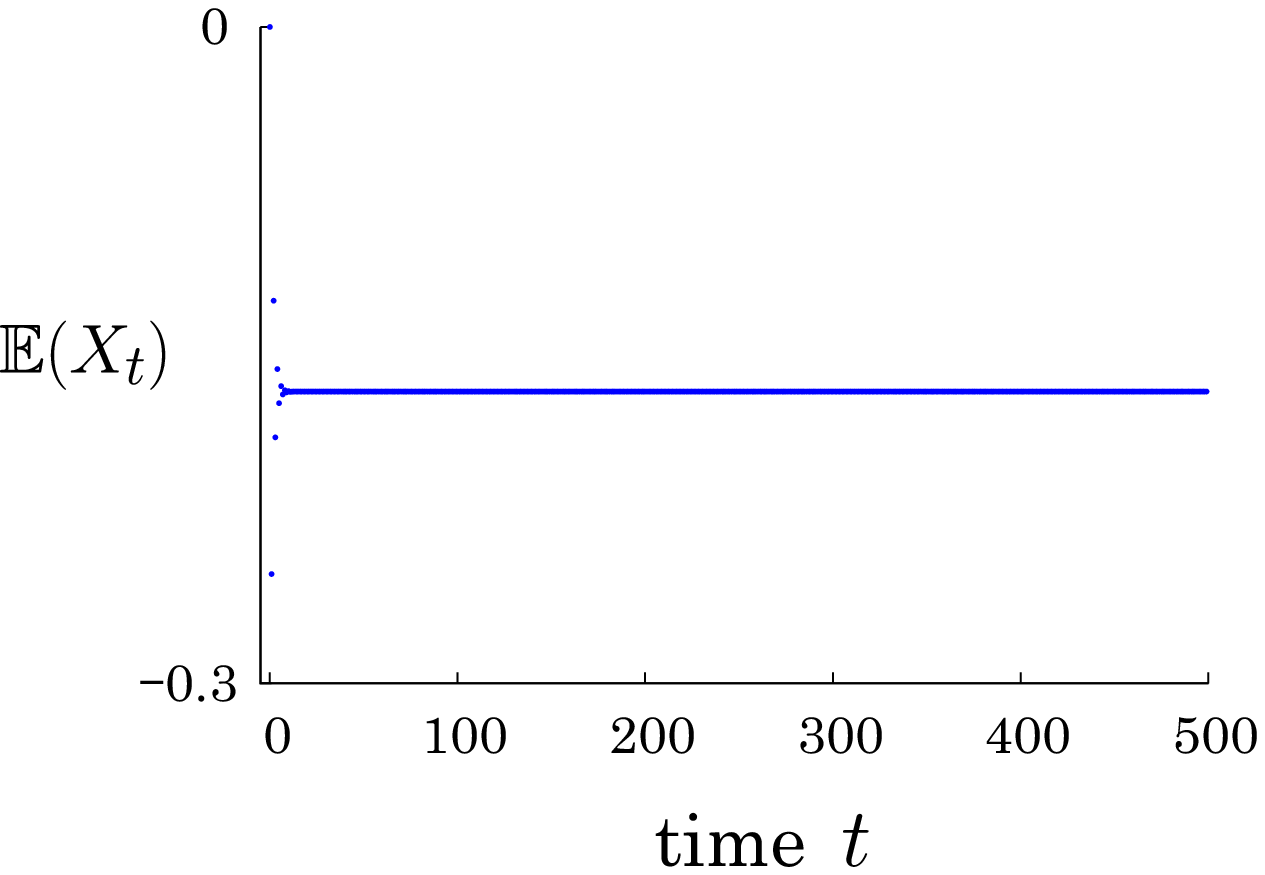}\\[2mm]
  (a) Numerical experiment
  \end{center}
 \end{minipage}
 \begin{minipage}{50mm}
  \begin{center}
   \includegraphics[scale=0.3]{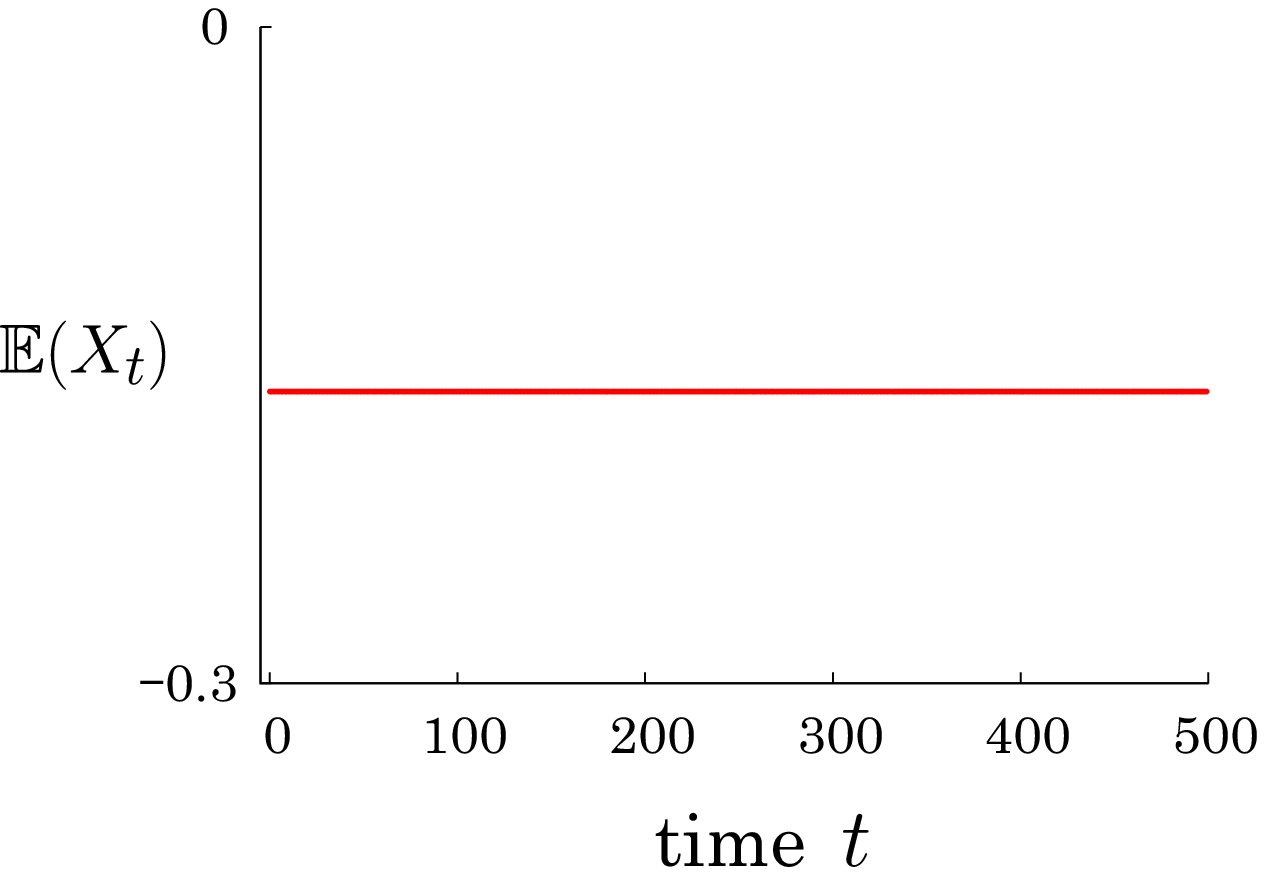}\\[2mm]
  (b) Approximation
  \end{center}
 \end{minipage}
\caption{(color figure online) $\theta_0=\pi/2,\,\theta_1=\pi/3$ : The $1$st moment $\mathbb{E}(X_t)$ of the open quantum walk with the initial state $D_0=\ket{0}\otimes (3/4\ket{0}\bra{0}+1/4\ket{1}\bra{1})$.}
\label{fig:10}
\end{center}
\end{figure}
\begin{figure}[h]
\begin{center}
 \begin{minipage}{50mm}
  \begin{center}
   \includegraphics[scale=0.3]{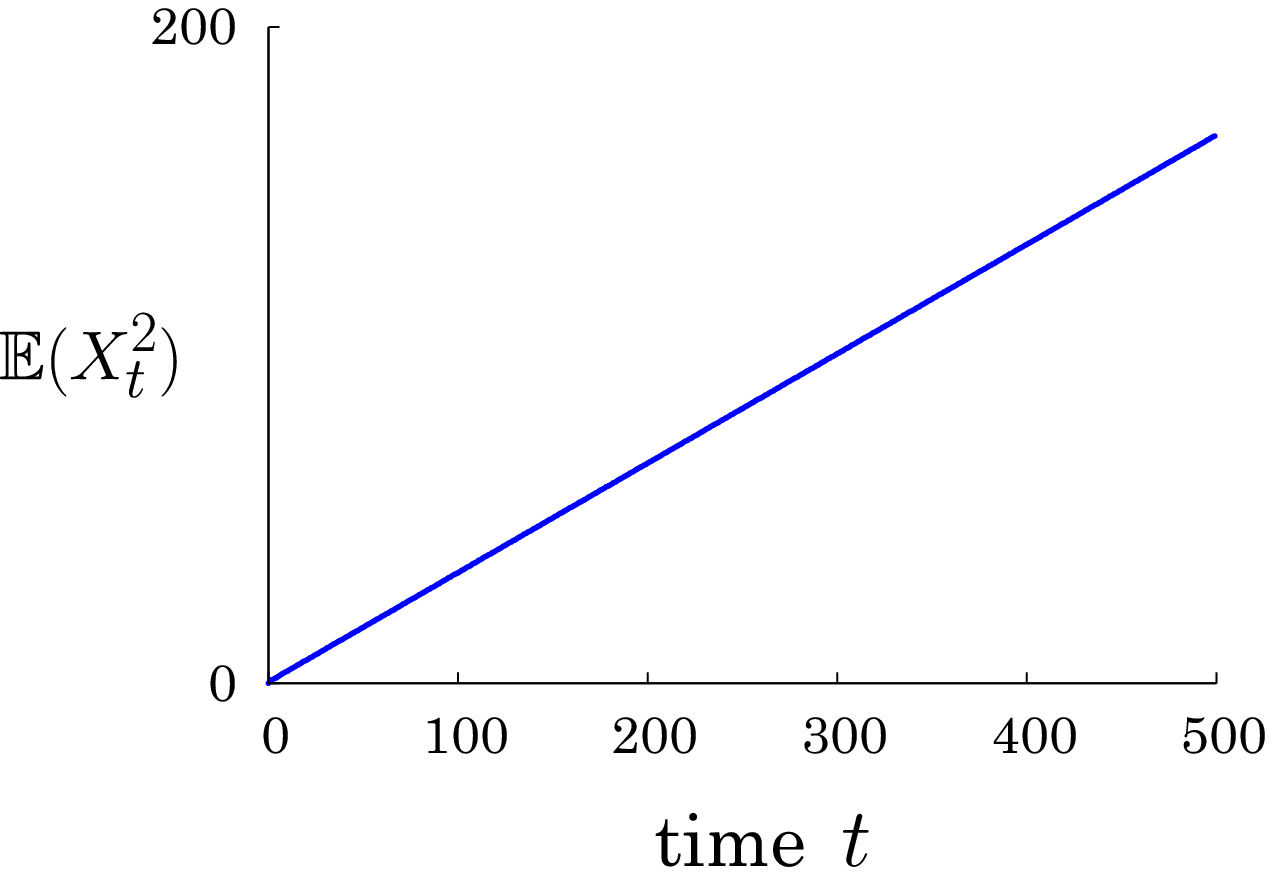}\\[2mm]
  (a) Numerical experiment
  \end{center}
 \end{minipage}
 \begin{minipage}{50mm}
  \begin{center}
   \includegraphics[scale=0.3]{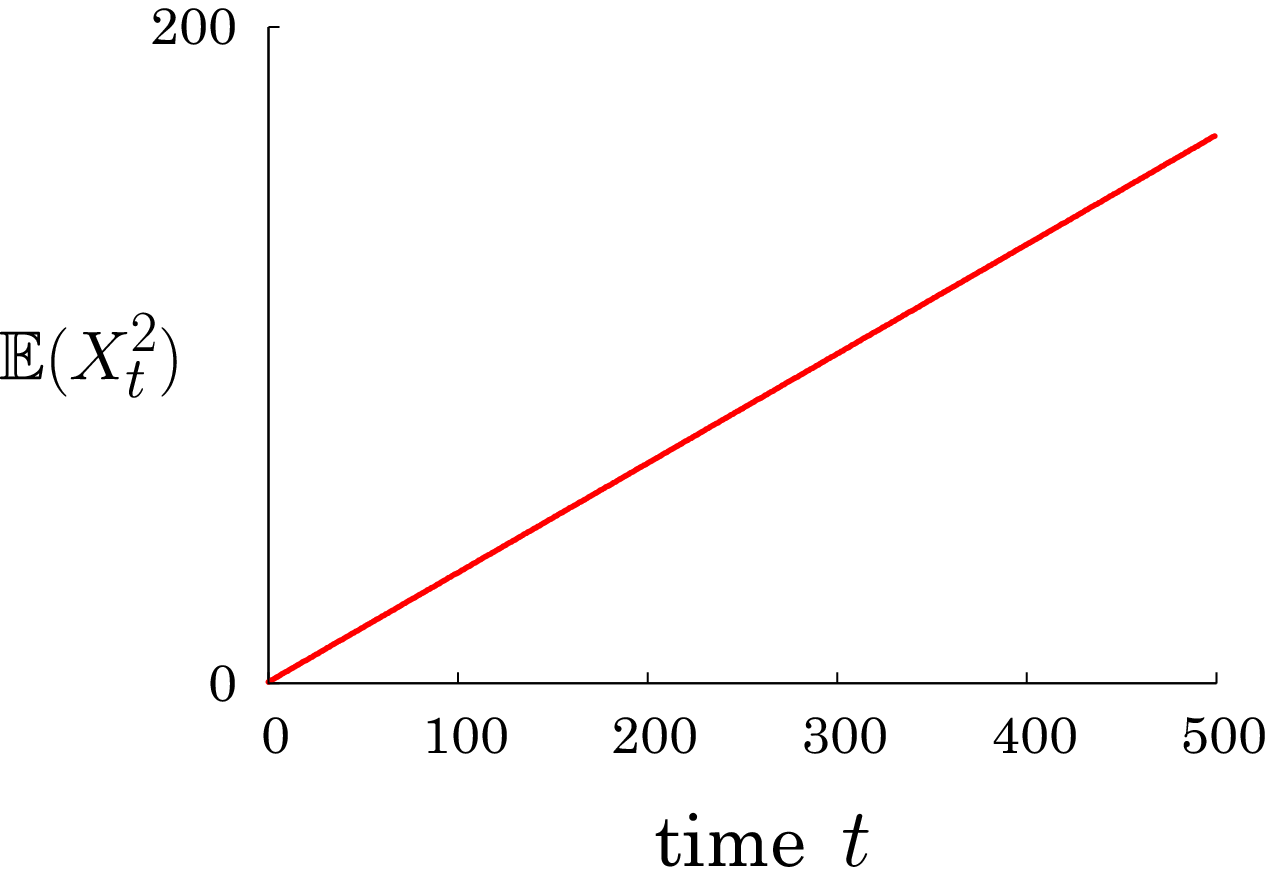}\\[2mm]
  (b) Approximation
  \end{center}
 \end{minipage}
\caption{(color figure online) $\theta_0=\pi/2,\,\theta_1=\pi/3$ : The $2$nd moment $\mathbb{E}(X_t^2)$ of the open quantum walk with the initial state $D_0=\ket{0}\otimes (3/4\ket{0}\bra{0}+1/4\ket{1}\bra{1})$.}
\label{fig:11}
\end{center}
\end{figure}
\begin{figure}[h]
\begin{center}
 \begin{minipage}{50mm}
  \begin{center}
   \includegraphics[scale=0.3]{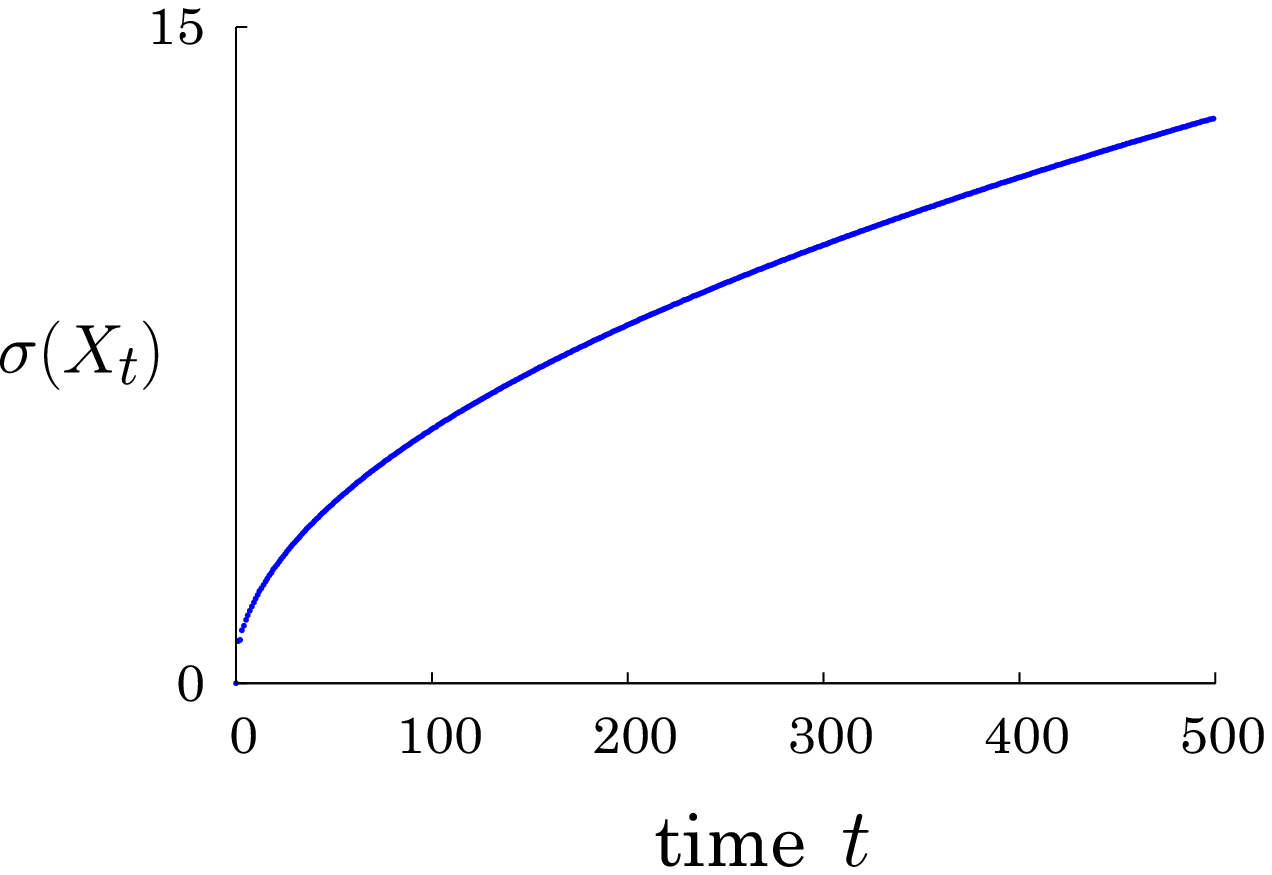}\\[2mm]
  (a) Numerical experiment
  \end{center}
 \end{minipage}
 \begin{minipage}{50mm}
  \begin{center}
   \includegraphics[scale=0.3]{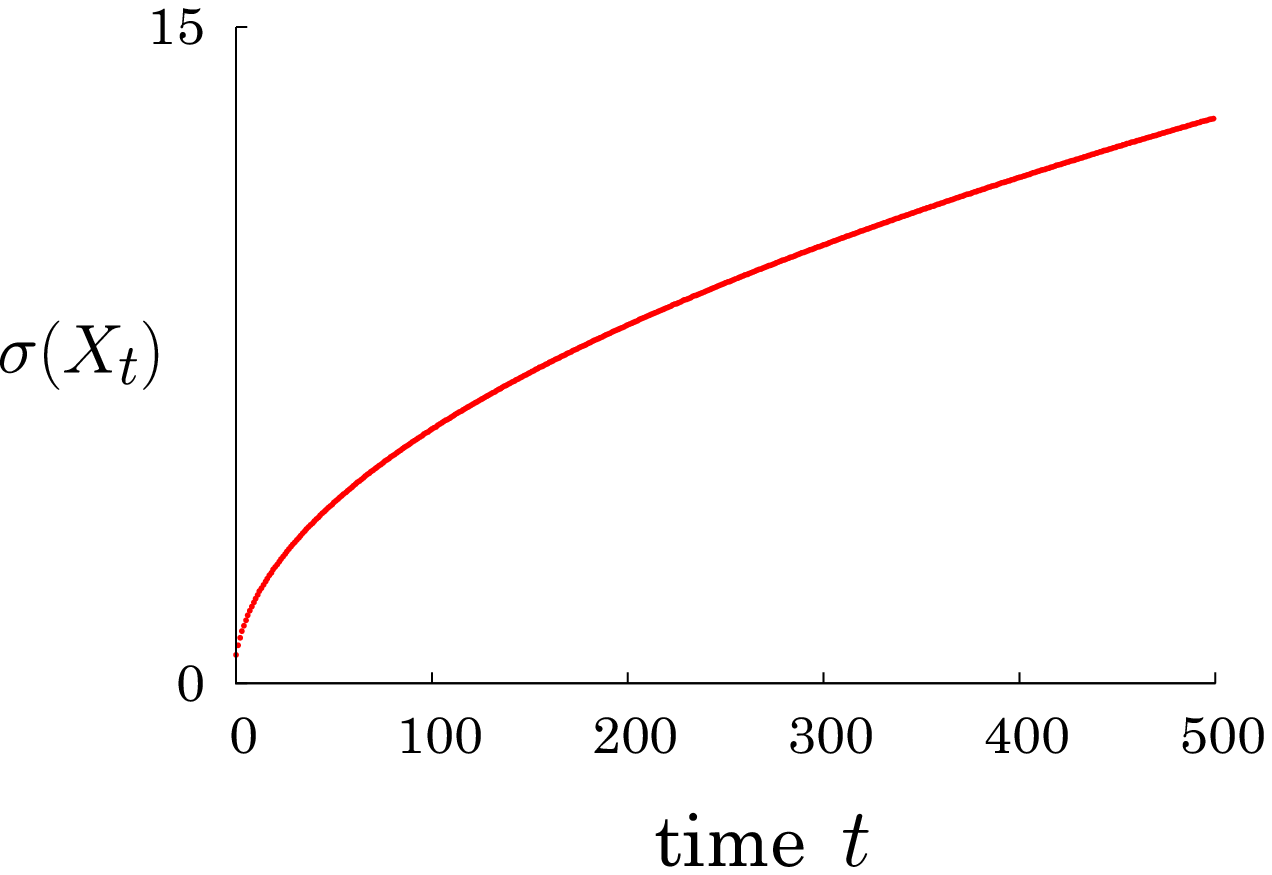}\\[2mm]
  (b) Approximation
  \end{center}
 \end{minipage}
\caption{(color figure online) $\theta_0=\pi/2,\,\theta_1=\pi/3$ : The standard deviation $\sigma(X_t)$ of the open quantum walk with the initial state $D_0=\ket{0}\otimes (3/4\ket{0}\bra{0}+1/4\ket{1}\bra{1})$.}
\label{fig:12}
\end{center}
\end{figure}

\clearpage

\begin{proof}{%
The operation $\hat{U}(k)$ becomes
\begin{equation}
 \hat{U}(k)=
 \begin{bmatrix}
  e^{-ik}c_1^2 & e^{-ik}s_1^2 & e^{-ik}c_1s_1 & e^{-ik}c_1s_1\\[2mm]
  e^{ik}s_1^2 & e^{ik}c_1^2 & -e^{ik}c_1s_1 & -e^{ik}c_1s_1\\[2mm]
  0 & 0 & 0 & 0\\[1mm]
  0 & 0 & 0 & 0
 \end{bmatrix},
\end{equation}
and the system of the open quantum walk at time $t$ results in
\begin{align}
 \ket{\hat\psi_t(k)}
 =&
  \begin{bmatrix}
   \begin{bmatrix}
    e^{-ik}c_1^2 & e^{-ik}s_1^2\\[1mm]
    e^{ik}s_1^2 & e^{ik}c_1^2
   \end{bmatrix}^{t}
   \begin{bmatrix}
    p\\1-p
   \end{bmatrix}\\
   0\\0
  \end{bmatrix}\nonumber\\
 =&
  \begin{bmatrix}
   \hat{V}(k)^t\,\ket{\phi}\\
   0\\0
  \end{bmatrix},\label{eq:201221-4}
\end{align}
where $\ket{\phi}=p\ket{0}+(1-p)\ket{1}$ and
\begin{equation}
 \hat{V}(k)=
 \begin{bmatrix}
  e^{-ik}c_1^2 & e^{-ik}s_1^2\\[1mm]
  e^{ik}s_1^2 & e^{ik}c_1^2
 \end{bmatrix}.
\end{equation}
Equation~\eqref{eq:201221-4} allows us to analyze the vector $\ket{\hat\psi_t(k)}\in\mathbb{C}^4$ in the reduced space $\mathbb{C}^2$.
The $2\times 2$ matrix $\hat{V}(k)$ holds two eigenvalues
\begin{equation}
 \nu_j(k)=c_1^2\cos k -(-1)^j\sqrt{s_1^4-c_1^4\sin^2k}\quad (j=1,2),
\end{equation}
from which $\nu_1(0)=1$ and $\nu_2(0)=c_1^2-s_1^2$.
A possible eigenvector associated to the eigenvalue $\nu_j(k)$, represented by $\ket{\xi_j(k)}$, is of the form
\begin{equation}
 \ket{\xi_j(k)}=
  \begin{bmatrix}
   -(-1)^j\sqrt{s_1^4-c_1^4\sin^2k}-i\,c_1^2\sin k\\[3mm]
   e^{ik}s_1^2
  \end{bmatrix}.
\end{equation}

Disassembling the reduced initial state in the eigenspace $\ket{\phi}=b_1(k)\ket{\xi_1(k)}+b_2(k)\ket{\xi_2(k)}$ and defining $z_j(k)=\bigl(\bra{0}+\bra{1}\bigr)\ket{\xi_j(k)}\,(j=1,2)$ again, we analyze the 1st and 2nd moments in a similar way to Eqs.~\eqref{eq:201221-2} and \eqref{eq:201221-3}.
We find
\begin{align}
 b_1(k)=& \frac{\braket{\xi_1(k)|\phi}\braket{\xi_2(k)|\xi_2(k)}-\braket{\xi_2(k)|\phi}\braket{\xi_1(k)|\xi_2(k)}}{\braket{\xi_1(k)|\xi_1(k)}\braket{\xi_2(k)|\xi_2(k)}-\braket{\xi_1(k)|\xi_2(k)}\braket{\xi_2(k)|\xi_1(k)}},\\
 b_2(k)=& \frac{\braket{\xi_2(k)|\phi}\braket{\xi_1(k)|\xi_1(k)}-\braket{\xi_1(k)|\phi}\braket{\xi_2(k)|\xi_1(k)}}{\braket{\xi_1(k)|\xi_1(k)}\braket{\xi_2(k)|\xi_2(k)}-\braket{\xi_1(k)|\xi_2(k)}\braket{\xi_2(k)|\xi_1(k)}},
\end{align}
where the denominator is organized to be a form,
\begin{align}
 & \braket{\xi_1(k)|\xi_1(k)}\braket{\xi_2(k)|\xi_2(k)}-\braket{\xi_1(k)|\xi_2(k)}\braket{\xi_2(k)|\xi_1(k)}\nonumber\\
 =& 4s_1^4(s_1^4-c_1^4\sin^2k).
\end{align}
Note that the vectors $\ket{\xi_1(0)}$ and $\ket{\xi_2(0)}$ are orthogonal to each other.
Since we see $s_1\neq 0$ from the conditions $c_0\neq s_0s_1$ and $c_0=0$, and $c_1\neq 0$ from $s_0c_1\neq 0$, the parameter $\theta_1$ is not fixed at $0, \pi/2, \pi, 3\pi/2$.
The value of $\nu_2(0)$ is, therefore, bounded,
\begin{equation}
 \bigl|\nu_2(0)\bigr|=\bigl|c_1^2-s_1^2\bigr|=\bigl|\cos2\theta_1\bigr|<1.
\end{equation}

One can approximately estimate the 1st and 2nd moments for large values of time $t$,
\begin{align}
 \mathbb{E}(X_t)
 =& \,i\,\Bigl(b_1(k)z_1(k)\Bigr)'\Big|_{k=0} + \nu_2(0)^t\,i\,\Bigl(b_2(k)z_2(k)\Bigr)'\Big|_{k=0}\nonumber\\[3mm]
 \sim &\, \,i\,\Bigl(b_1(k)z_1(k)\Bigr)'\Big|_{k=0}\nonumber\\[3mm]
 =&\,\frac{(2p-1)(1-2s_1^2)}{2s_1^2},\\[3mm]
 \mathbb{E}(X_t^2)
 =& \,t\,\Bigl\{-\nu''_1(0)b_1(0)z_1(0)-\nu_2(0)^{t-1}\,\nu''_2(0)b_2(0)z_2(0)\Bigr\}\nonumber\\
 & -\Bigl(b_1(k)z_1(k)\Bigr)''\Big|_{k=0} - \nu_2(0)^t\,\Bigl(b_2(k)z_2(k)\Bigr)''\Big|_{k=0}\nonumber\\[3mm]
 \sim & \,-t\,\nu''_1(0)b_1(0)z_1(0)-\Bigl(b_1(k)z_1(k)\Bigr)''\Big|_{k=0}\nonumber\\[3mm]
 =&\, \frac{c_1^2}{s_1^2}\,t - \frac{1-2s_1^2}{2s_1^4}.
\end{align}
\vspace{0mm}
}\end{proof}

\clearpage

	\item Case: $c_0\neq 0$

	      Although two probability distributions seem to be similar in Fig.~\ref{fig:13}, the values of their 1st moments $\mathbb{E}(X_{500})$ are different from each other. Numerical experiments compute the values, $\mathbb{E}(X_{500})=0$ in Fig.~\ref{fig:13}-(a) and $\mathbb{E}(X_{500})=0.333\cdots$ in Fig.~\ref{fig:13}-(b). 
\begin{figure}[h]
\begin{center}
 \begin{minipage}{50mm}
  \begin{center}
   \includegraphics[scale=0.4]{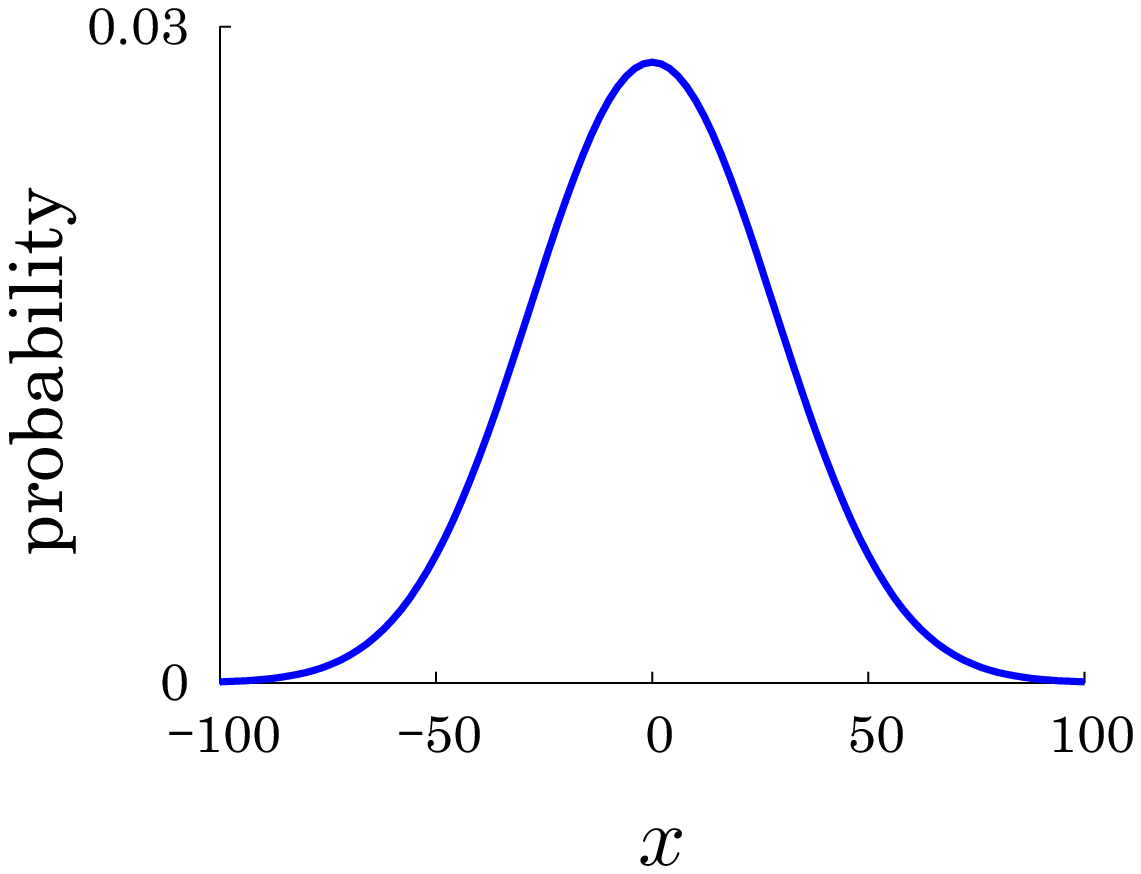}\\[2mm]
  (a) $p=1/2$
  \end{center}
 \end{minipage}
 \begin{minipage}{50mm}
  \begin{center}
   \includegraphics[scale=0.4]{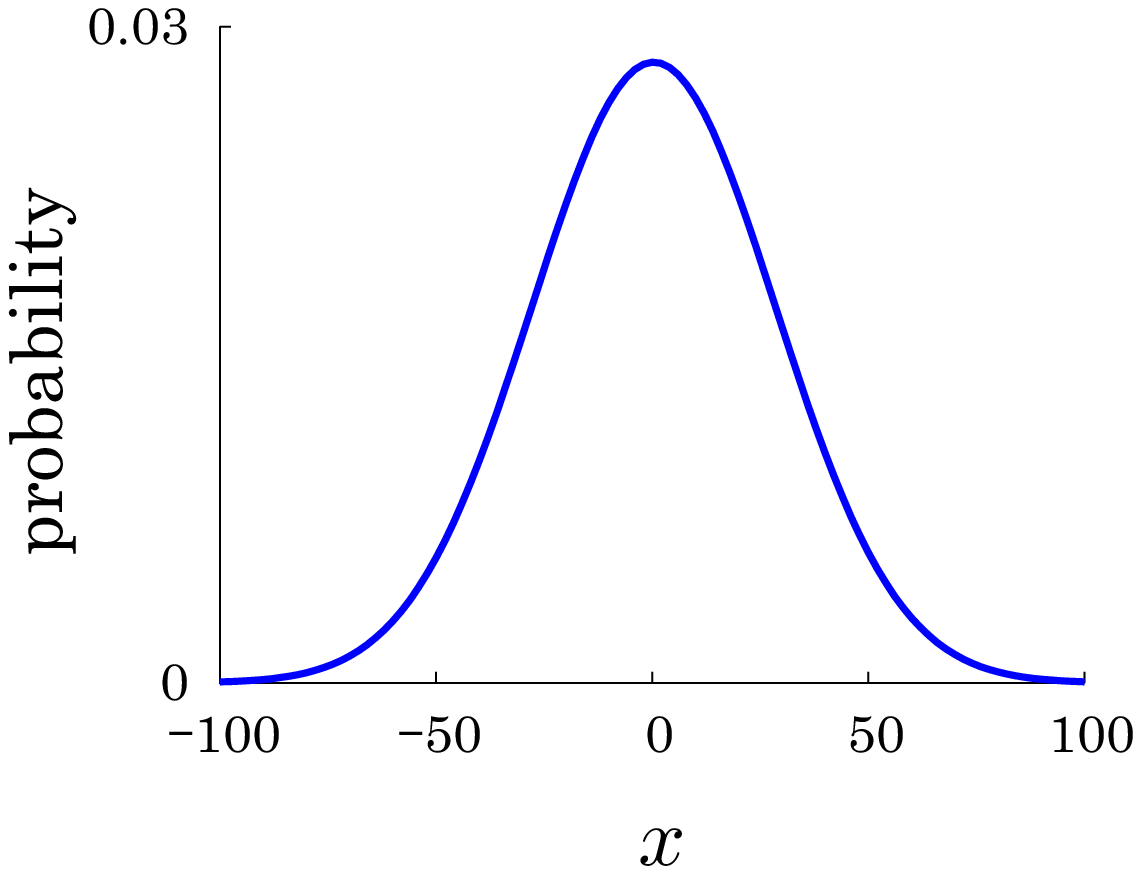}\\[2mm]
  (b) $p=3/4$
  \end{center}
 \end{minipage}
\vspace{5mm}
\caption{(color figure online) $\theta_0=\pi/6,\,\theta_1=\pi/3$ : Probability distribution of the open quantum walk at time $t=500$ with the initial state $D_0=\ket{0}\otimes (p\ket{0}\bra{0}+(1-p)\ket{1}\bra{1})$.}
\label{fig:13}
\end{center}
\end{figure}

	      One can prove that the 1st moment converges to a value as time $t\to\infty$, and have an approximation to the 2nd moment for large values of time $t$,
	      \begin{align}
	       \mathbb{E}(X_t)= & O(1),\\[3mm]
	       \mathbb{E}(X_t^2)\sim & \frac{s_0^2\left\{c_1^2+4c_0^2s_1^2+4c_0s_0s_1(c_1^2-2c_0^2s_1^2)\right\}}{(c_0-s_0s_1)^2}\,t,\\[3mm]
	       \sigma(X_t)\sim & \frac{|s_0|\sqrt{c_1^2+4c_0^2s_1^2+4c_0s_0s_1(c_1^2-2c_0^2s_1^2)}}{|c_0-s_0s_1|}\sqrt{t}.
	      \end{align}
	      The convergence of the 1st moment is observed in a numerical experiment, as Fig.~\ref{fig:14} shows.
	      The approximations to the 2nd moment and the standard deviation are compared to the actual values in Figs.~\ref{fig:15} and \ref{fig:16}.
\clearpage

\begin{figure}[h]
\begin{center}
 \includegraphics[scale=0.3]{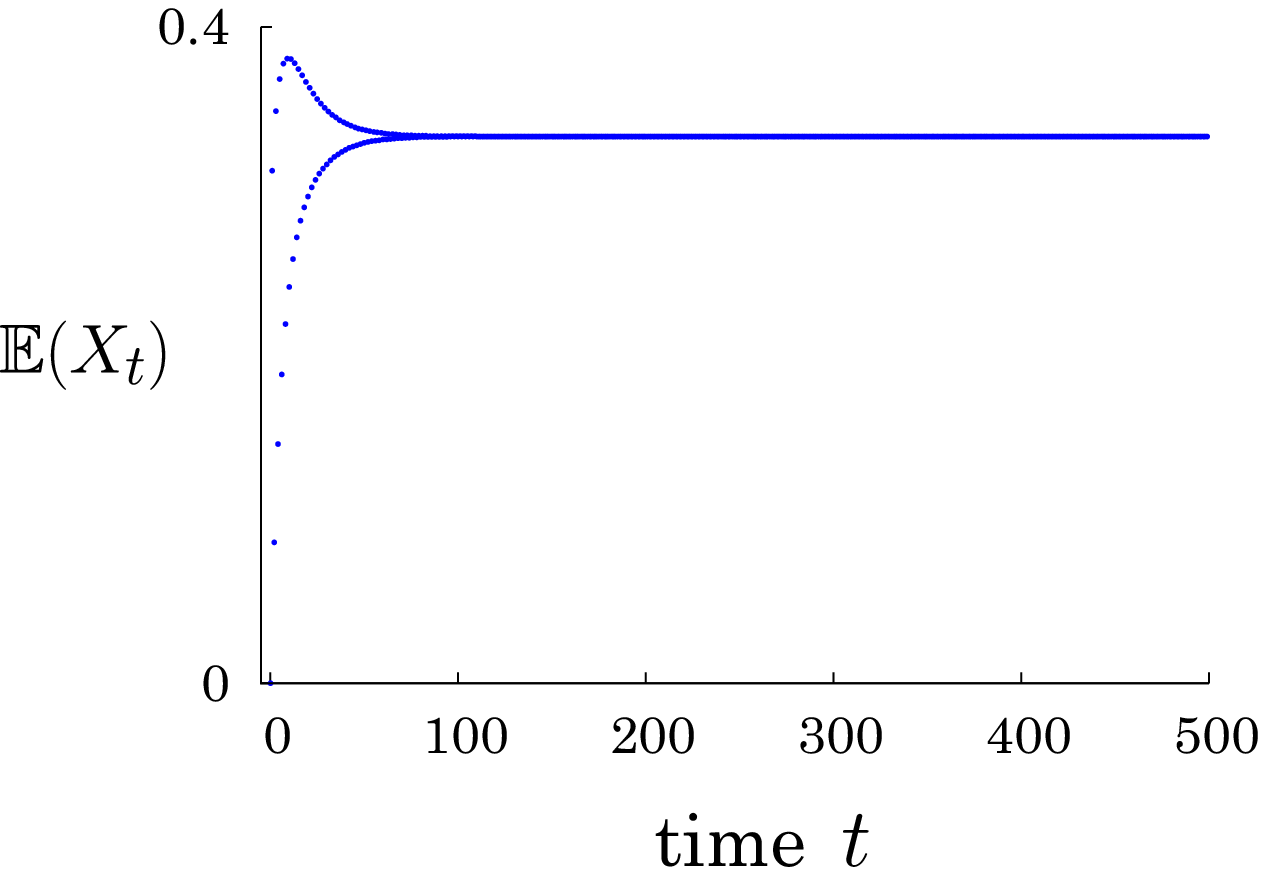}\\[2mm]
 Numerical experiment
\caption{(color figure online) $\theta_0=\pi/6,\,\theta_1=\pi/3$ : The $1$st moment $\mathbb{E}(X_t)$ of the open quantum walk with the initial state $D_0=\ket{0}\otimes (3/4\ket{0}\bra{0}+1/4\ket{1}\bra{1})$.}
\label{fig:14}
\end{center}
\end{figure}
\begin{figure}[h]
\begin{center}
 \begin{minipage}{50mm}
  \begin{center}
   \includegraphics[scale=0.3]{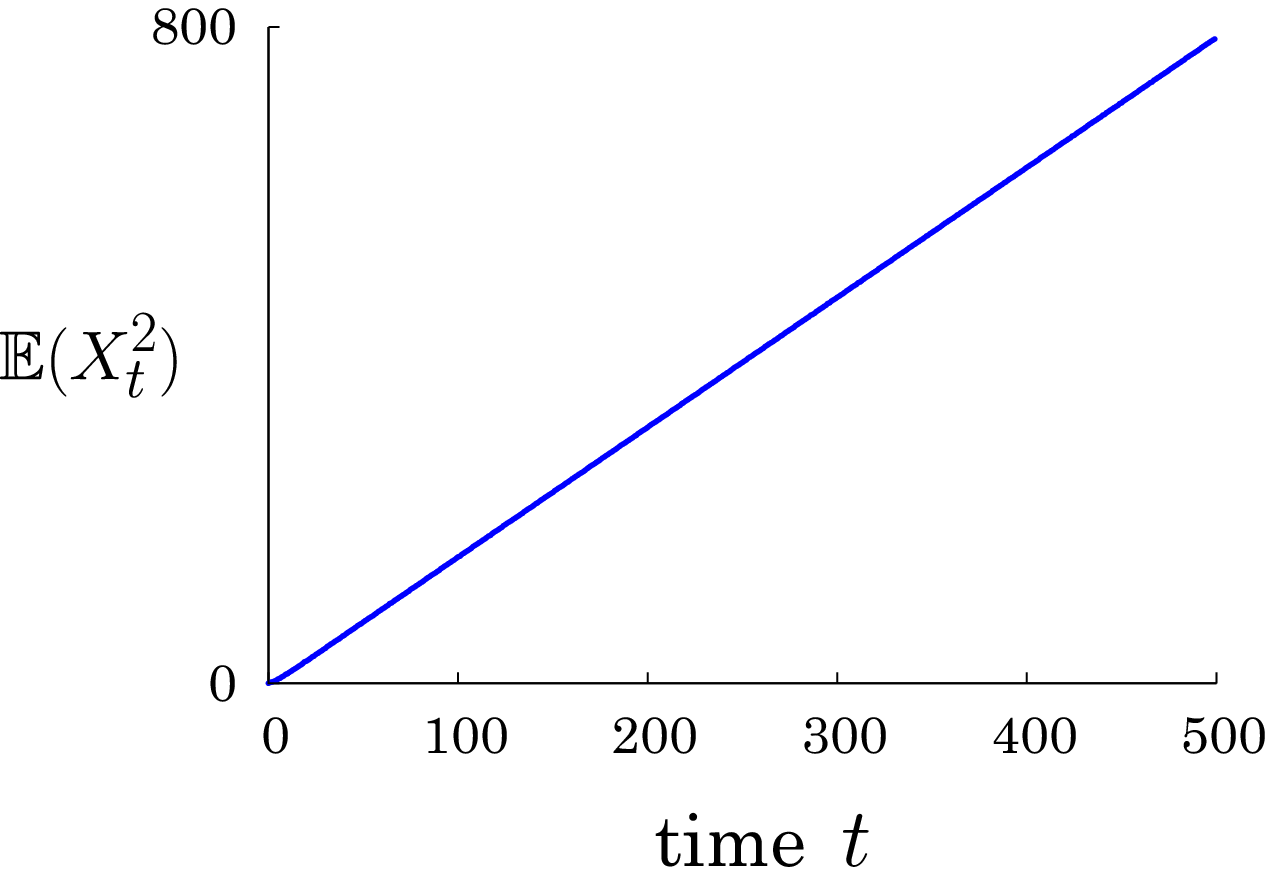}\\[2mm]
  (a) Numerical experiment
  \end{center}
 \end{minipage}
 \begin{minipage}{50mm}
  \begin{center}
   \includegraphics[scale=0.3]{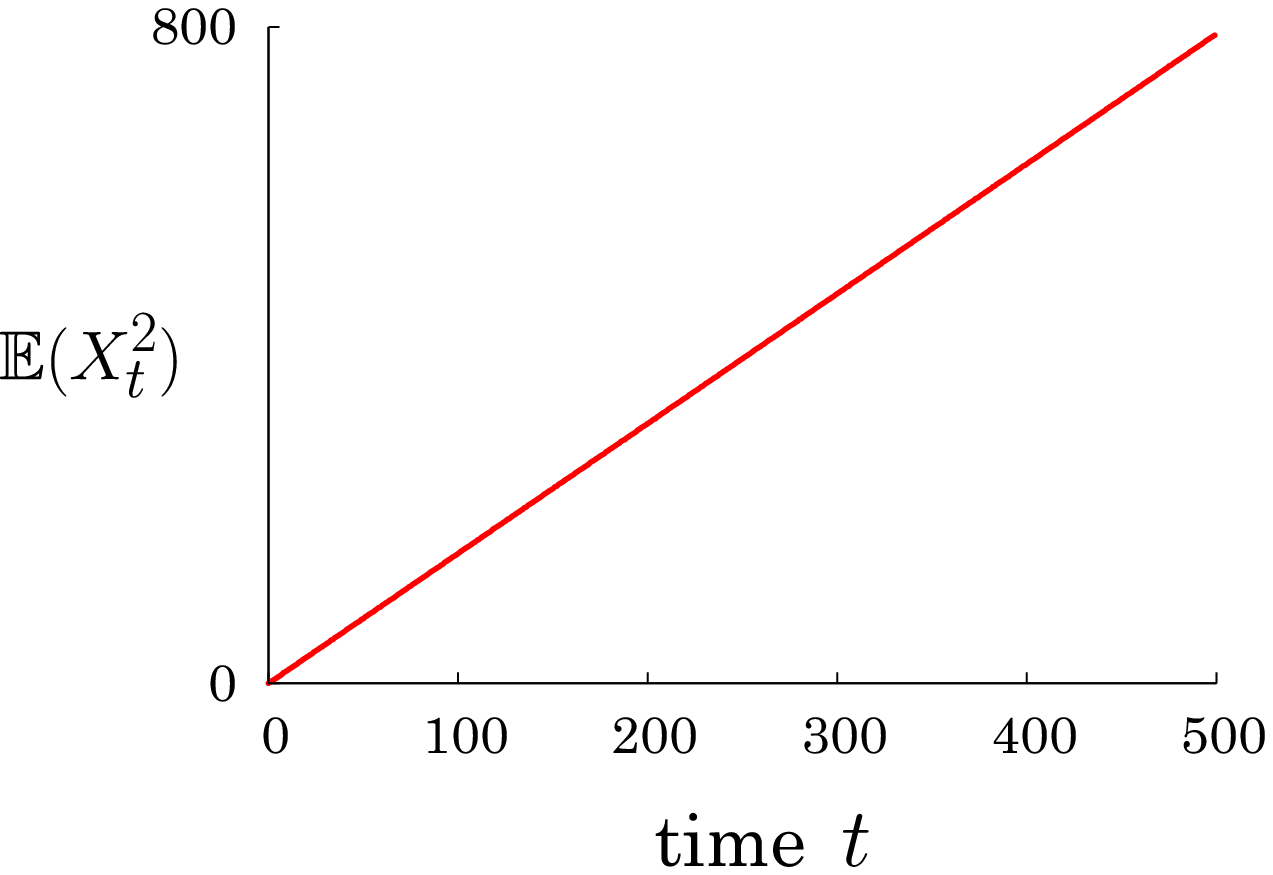}\\[2mm]
  (b) Approximation
  \end{center}
 \end{minipage}
\caption{(color figure online) $\theta_0=\pi/6,\,\theta_1=\pi/3$ : The $2$nd moment $\mathbb{E}(X_t^2)$ of the open quantum walk with the initial state $D_0=\ket{0}\otimes (3/4\ket{0}\bra{0}+1/4\ket{1}\bra{1})$.}
\label{fig:15}
\end{center}
\end{figure}
\begin{figure}[h]
\begin{center}
 \begin{minipage}{50mm}
  \begin{center}
   \includegraphics[scale=0.3]{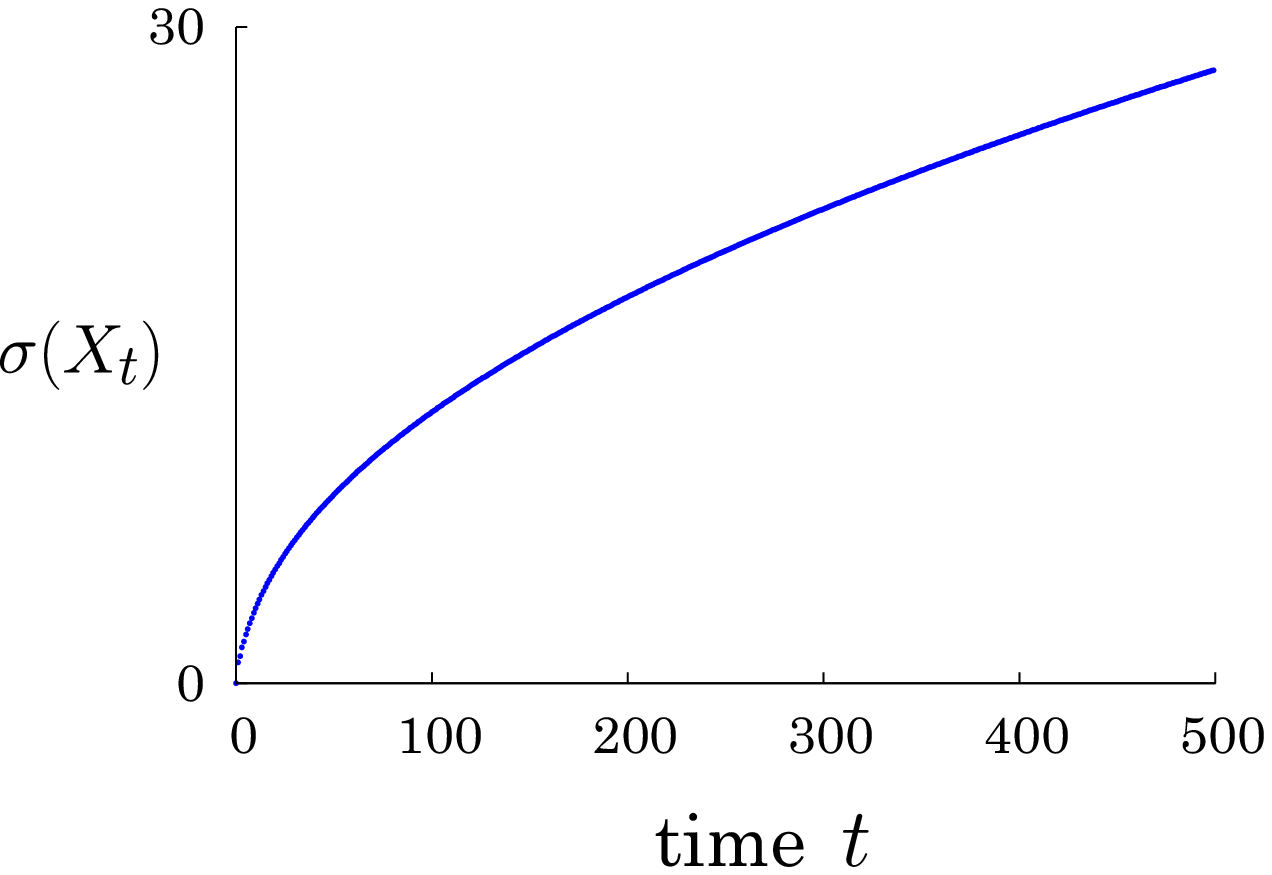}\\[2mm]
  (a) Numerical experiment
  \end{center}
 \end{minipage}
 \begin{minipage}{50mm}
  \begin{center}
   \includegraphics[scale=0.3]{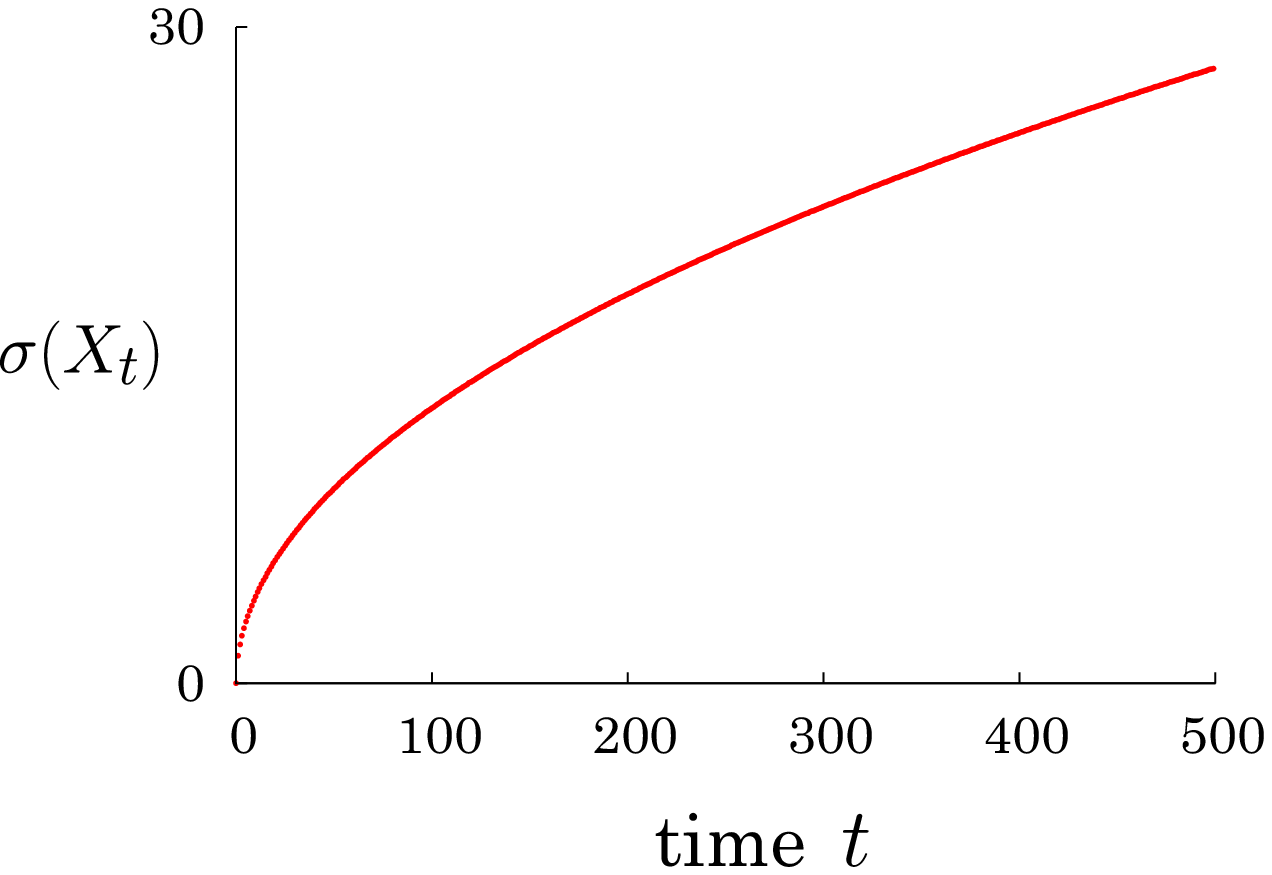}\\[2mm]
  (b) Approximation
  \end{center}
 \end{minipage}
\caption{(color figure online) $\theta_0=\pi/6,\,\theta_1=\pi/3$ : The standard deviation $\sigma(X_t)$ of the open quantum walk with the initial state $D_0=\ket{0}\otimes (3/4\ket{0}\bra{0}+1/4\ket{1}\bra{1})$.}
\label{fig:16}
\end{center}
\end{figure}

       \end{enumerate}
\end{enumerate}

\clearpage

\begin{proof}{%
Let $I_4$ be the $4\times 4$ identity matrix, 
\begin{equation}
 I_4=\ket{00}\bra{00}+\ket{01}\bra{01}+\ket{10}\bra{10}+\ket{11}\bra{11}.
\end{equation}
Then, the characteristic polynomial of the matrix $\hat{U}(k)$ is solved to two factors,
\begin{align}
 & \det\left(\hat{U}(k)-\lambda\, I_4\right)\nonumber\\
 =& (\lambda+2c_0s_0s_1\cos k)\biggl[\lambda^3 -2s_0(s_0c_1^2+c_0s_1)(\cos k)\,\lambda^2\nonumber\\
 & +\Bigl\{4c_0s_0^2s_1(s_0c_1^2+c_0s_1)\sin^2 k +2s_0^2c_1^2-1\Bigr\}\lambda\nonumber\\
 & +2c_0s_0s_1(1-4c_0^2s_0^2s_1^2\sin^2 k)\cos k\biggr]\nonumber\\
 =& (\lambda+2c_0s_0s_1\cos k)\,G(\lambda),
\end{align}
in which we have defined a degree 3 polynomial function of $\lambda$,
\begin{align}
 G(\lambda)
 =& \lambda^3 -2s_0(s_0c_1^2+c_0s_1)(\cos k)\,\lambda^2\nonumber\\
 & +\Bigl\{4c_0s_0^2s_1(s_0c_1^2+c_0s_1)\sin^2 k +2s_0^2c_1^2-1\Bigr\}\lambda\nonumber\\
 & +2c_0s_0s_1(1-4c_0^2s_0^2s_1^2\sin^2 k)\cos k.
\end{align}

We denote the four eigenvalues of $\hat{U}(k)$ by $\lambda_j(k)\,(j=1,2,3,4)$, three of which satisfy $G(\lambda_j(k))=0\,(j=1,2,3)$ and one of which is $\lambda_4(k)=-2c_0s_0s_1\cos k$.
With three functions of $k\in[-\pi, \pi)$,
\begin{equation}
 g_j(k)=G(\lambda_j(k))\quad (j=1,2,3),
\end{equation}
since we have
\begin{equation}
 g_j(0)=(\lambda_j(0)-1)\Bigl[\lambda_j(0)^2-\left\{s_0^2c_1^2-(c_0-s_0s_1)^2\right\}\lambda_j(0)-2c_0s_0s_1\Bigr],
\end{equation}
the eigenvalue $\lambda_1(k)$ is allowed to be the value such that $G(\lambda_1(k))=0$ and $\lambda_1(0)=1$, and the eigenvalue $\lambda_j(k)\,(j\in\left\{2,3\right\})$ is such that $G(\lambda_j(k))=0$ and
\begin{equation}
 \lambda_j(0)=\frac{1}{2}\left[s_0^2c_1^2-(c_0-s_0s_1)^2+(-1)^j\sqrt{\Bigl\{s_0^2c_1^2-(c_0-s_0s_1)^2\Bigr\}^2+8c_0s_0s_1}\,\right],
\end{equation}
where the values of $\lambda_j(0)\,(j=2,3)$ are bounded, that is $\bigl|\lambda_j(0)\bigr|<1\,(j=2,3)$, because the quadratic equation regarding $\lambda$,
\begin{equation}
 \lambda^2-\left\{s_0^2c_1^2-(c_0-s_0s_1)^2\right\}\lambda-2c_0s_0s_1=0,
\end{equation}
holds two solutions such that $|\lambda|<1$ under the conditions $c_0\neq s_0s_1$ and $s_0c_1\neq 0$.
The value of $\bigl|\lambda_4(0)\bigr|$ is also bounded to be less than 1 because of $|2c_0s_0s_1|<1$ under the condition $s_0c_1\neq 0$.

A possible eigenvector associated to the eigenvalue $\lambda_j(k)\,(j\in\left\{1,2,3,4\right\})$, represented by $\ket{v_j(k)}$, is computed and they are in the following forms,
\begin{align}
 \ket{v_1(k)}=&
 \begin{bmatrix}
  1\\1\\0\\0
 \end{bmatrix},\\[3mm]
 \ket{v_2(k)}=&
 \begin{bmatrix}
  -c_0^2-2c_0s_0s_1+s_0^2(c_1^2-s_1^2)+\sqrt{J(\theta_0,\theta_1)}\,\,\\[2mm]
  c_0^2+2c_0s_0s_1-s_0^2(c_1^2-s_1^2)-\sqrt{J(\theta_0,\theta_1)}\\[2mm]
  4c_0s_0c_1\\[1mm]
  4c_0s_0c_1
 \end{bmatrix},\\[3mm]
 \ket{v_3(k)}=&
 \begin{bmatrix}
  -c_0^2-2c_0s_0s_1+s_0^2(c_1^2-s_1^2)-\sqrt{J(\theta_0,\theta_1)}\,\,\\[2mm]
  c_0^2+2c_0s_0s_1-s_0^2(c_1^2-s_1^2)+\sqrt{J(\theta_0,\theta_1)}\\[2mm]
  4c_0s_0c_1\\[1mm]
  4c_0s_0c_1
 \end{bmatrix},\\[3mm]
 \ket{v_4(k)}=&
 \begin{bmatrix}
  0\\0\\-1\\1
 \end{bmatrix},
\end{align}
where $J(\theta_0,\theta_1)=\left\{s_0^2c_1^2-(c_0-s_0s_1)^2\right\}^2+8c_0s_0s_1$.
Since the vector $\ket{v_1(0)}$ is orthogonal to $\ket{v_2(0)}$, $\ket{v_3(0)}$, and $\ket{v_4(0)}$,
the orthogonality derives
\begin{equation}
 a_1(0)=\frac{\braket{v_1(0)|\hat\psi_0(0)}}{\braket{v_1(0)|v_1(0)}}=\frac{1}{2}.
\end{equation}

Considering $\lambda_1(0)=1$, one can prove $\lambda'_1(0)=0$ from $G(\lambda_1(k))=0$,
\begin{equation}
 \frac{d}{dk}G(\lambda_1(k))\Big|_{k=0}=0\quad \Leftrightarrow\quad g'_1(0)=0\quad \Leftrightarrow\quad 2(c_0-s_0s_1)^2\lambda'_1(0)=0.\label{eq:201221-5}
\end{equation}
The condition $c_0\neq s_0s_1$ was given in this section (Sect.~\ref{subsec:c0!=s0s1}) so that we find $\lambda'_1(0)=0$ from Eq.~\eqref{eq:201221-5}. 
With $\lambda_1(0)=1$ and $\lambda'_1(0)=0$, a similar way finds the 2nd derivative of $\lambda_1(k)$ at the point $k=0$,
\begin{equation}
 \lambda''_1(0)=-\,\frac{s_0^2\left\{c_1^2+4c_0^2s_1^2+4c_0s_0s_1(c_1^2-2c_0^2s_1^2)\right\}}{(c_0-s_0s_1)^2},
\end{equation}
which comes from
\begin{equation}
 \frac{d^2}{dk^2}G(\lambda_1(k))\Big|_{k=0}=0\quad \Leftrightarrow\quad g''_1(0)=0.
\end{equation}

Reminding $\lambda_1(0)=1$ and $\bigl|\lambda_j(0)\bigr|<1\,(j=2,3,4)$, we reach approximations for large values of time $t$,
\begin{align}
 \mathbb{E}(X_t)\sim\, & \,i\,\Bigl(a_1(k)w_1(k)\Bigr)'\Big|_{k=0}\,=\,O(1),\\[3mm]
 \mathbb{E}(X_t^2)\sim\, &\,
 \frac{s_0^2\left\{c_1^2+4c_0^2s_1^2+4c_0s_0s_1(c_1^2-2c_0^2s_1^2)\right\}}{(c_0-s_0s_1)^2}\,t
 -\Bigl(a_1(k)w_1(k)\Bigr)''\Big|_{k=0}\nonumber\\
 \sim\,& \,\frac{s_0^2\left\{c_1^2+4c_0^2s_1^2+4c_0s_0s_1(c_1^2-2c_0^2s_1^2)\right\}}{(c_0-s_0s_1)^2}\,t.
\end{align}
\vspace{0mm}
}\end{proof}

\clearpage

The analysis, which we have demonstrated, gives a limit to a rescaled standard deviation as $t\to\infty$, and they are combined as a theorem.
\begin{thm}
The open quantum walk launches with the initial state $D_0=\ket{0}\otimes \bigl(p\ket{0}\bra{0}+(1-p)\ket{1}\bra{1}\,\bigr)\,(p\in [0,1]\,)$ and its system is updated by the matrices $P(\theta_0,\theta_1)$ and $Q(\theta_0,\theta_1)$ in Eqs.~\eqref{eq:matrix_P} and \eqref{eq:matrix_Q}.
The standard deviation $\sigma(X_t)$ increases in a different order of time $t$, depending on the values of parameters $\theta_0$ and $\theta_1$.
The standard deviation suitably rescaled by time $t$, converges to a value below.
 \begin{enumerate}
  \item $c_0=s_0s_1$

	\begin{enumerate}
	 \item $c_1\neq 0$
	       \begin{equation}
		\lim_{t\to\infty}\frac{\sigma(X_t)}{t} = s_0^2|c_1|\sqrt{1+3s_1^2-(2p-1)^2c_1^2}\label{eq:limit_ballistic}
	       \end{equation}

	 \item $c_1=0$
	       \begin{equation}
	 \lim_{t\to\infty}\frac{\sigma(X_t)}{\sqrt{t}} = 1\label{eq:limit_diffusive2}
	\end{equation}
	\end{enumerate}

  \item $c_0\neq s_0s_1$
	\begin{equation}
	 \lim_{t\to\infty}\frac{\sigma(X_t)}{\sqrt{t}} = \frac{|s_0|\sqrt{c_1^2+4c_0^2s_1^2+4c_0s_0s_1(c_1^2-2c_0^2s_1^2)}}{|c_0-s_0s_1|}\label{eq:limit_diffusive}
	\end{equation}
	
 \end{enumerate}
\end{thm}

\clearpage

\begin{figure}[h]
\begin{center}
 \begin{minipage}{50mm}
  \begin{center}
   \includegraphics[scale=0.4]{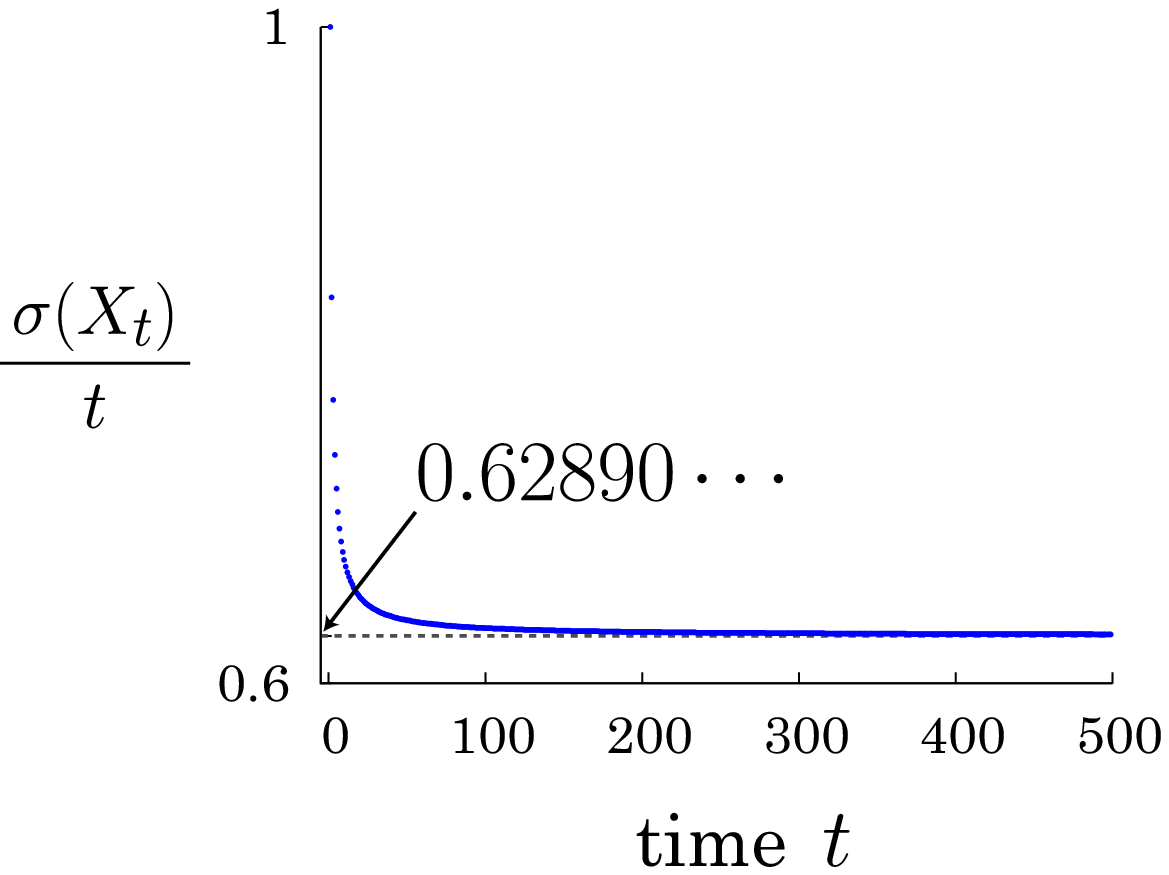}\\[2mm]
  (a) $p=1/2$
  \end{center}
 \end{minipage}
 \begin{minipage}{50mm}
  \begin{center}
   \includegraphics[scale=0.4]{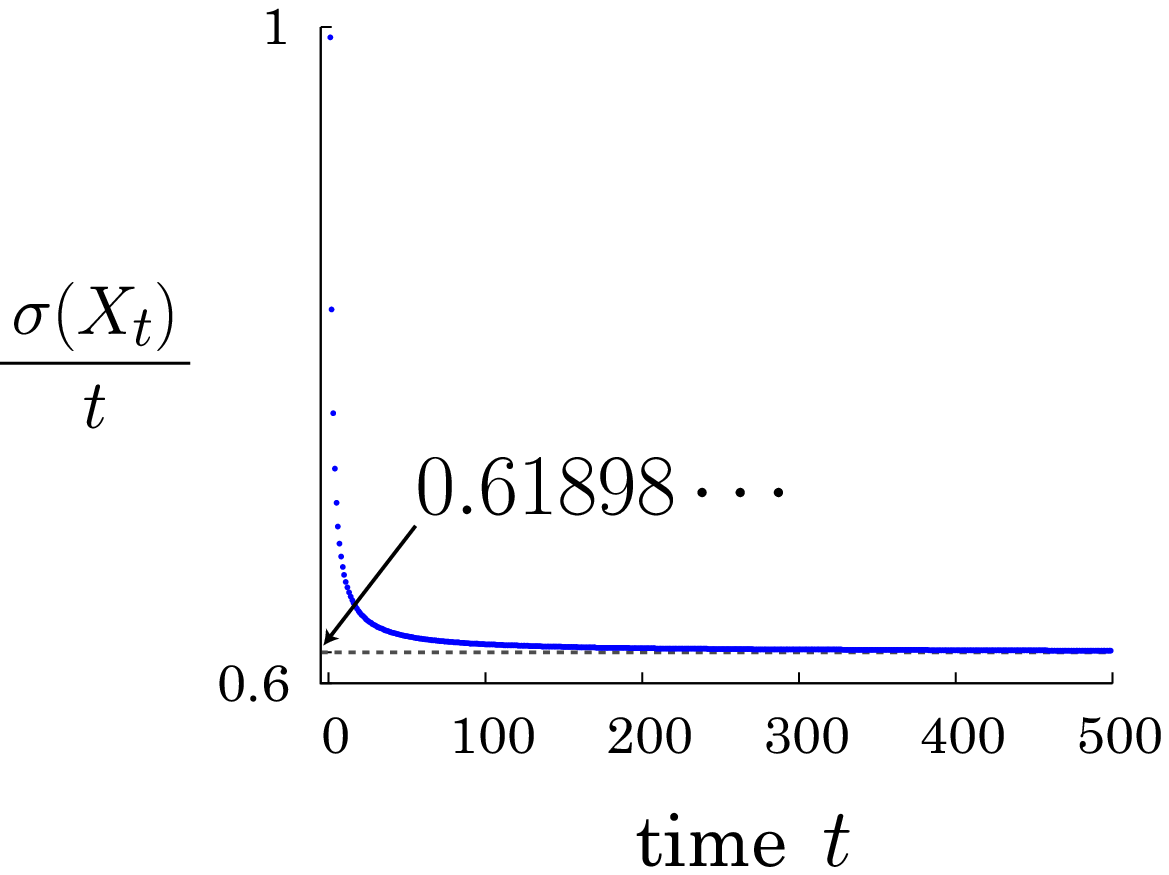}\\[2mm]
  (b) $p=3/4$
  \end{center}
 \end{minipage}
\caption{(color figure online) $\theta_0=2\pi/7,\,\theta_1=\arcsin(c_0/s_0)$ : The time-rescaled standard deviation $\sigma(X_t)/t$ converges to a constant as $t\to\infty$. The open quantum walk launches with the initial state $D_0=\ket{0}\otimes (p\ket{0}\bra{0}+(1-p)\ket{1}\bra{1})$.}
\label{fig:17}
\end{center}
\end{figure}

\begin{figure}[h]
\begin{center}
 \begin{minipage}{50mm}
  \begin{center}
   \includegraphics[scale=0.4]{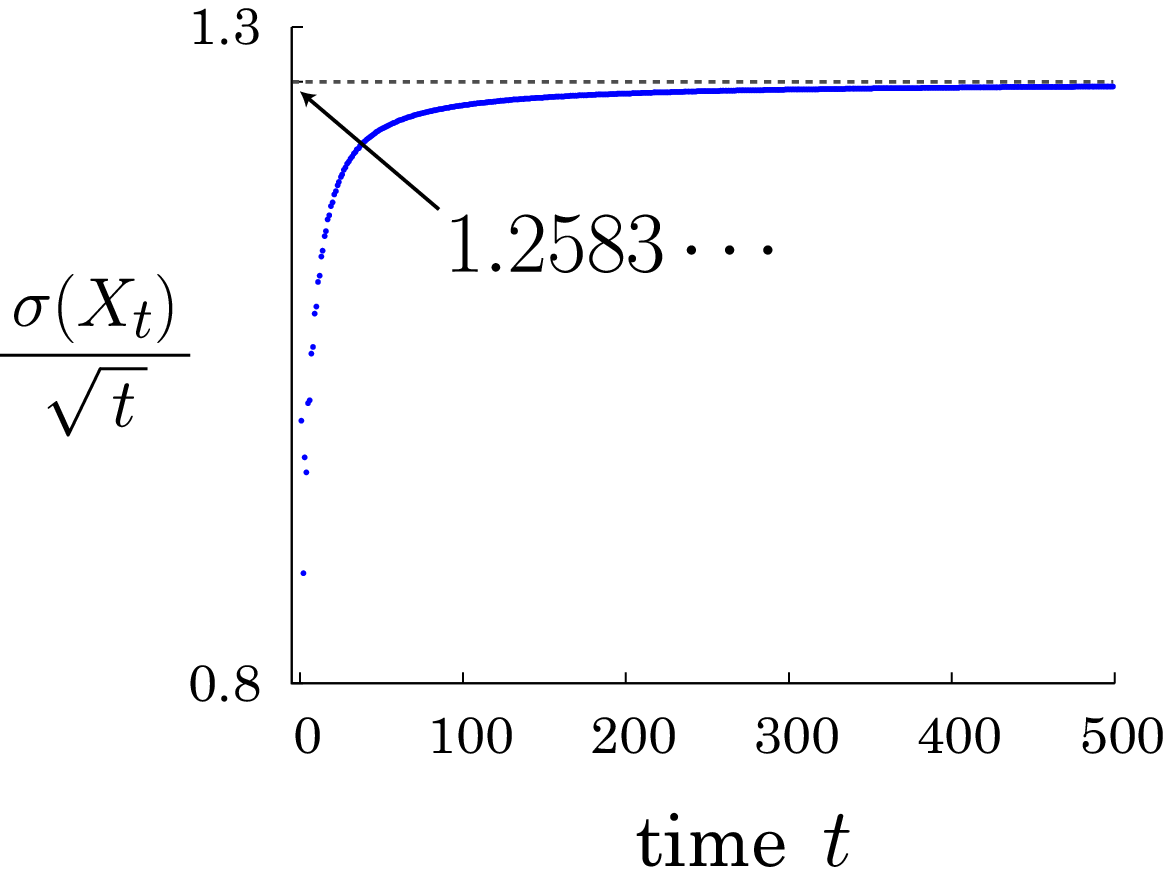}\\[2mm]
  (a) $p=1/2$
  \end{center}
 \end{minipage}
 \begin{minipage}{50mm}
  \begin{center}
   \includegraphics[scale=0.4]{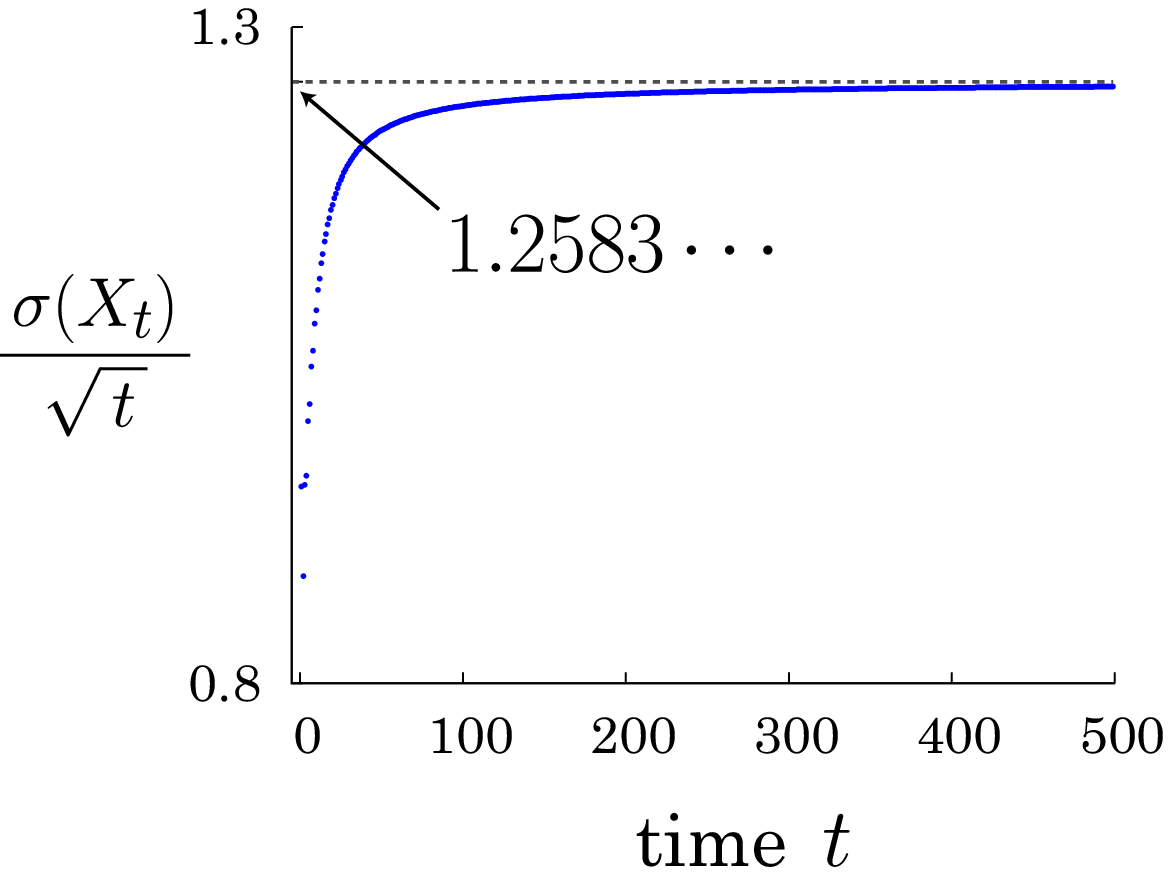}\\[2mm]
  (b) $p=3/4$
  \end{center}
 \end{minipage}
\caption{(color figure online) $\theta_0=\pi/6,\,\theta_1=\pi/3$ : The time-rescaled standard deviation $\sigma(X_t)/\sqrt{t}$ converges to a constant as $t\to\infty$. The open quantum walk launches with the initial state $D_0=\ket{0}\otimes (p\ket{0}\bra{0}+(1-p)\ket{1}\bra{1})$.}
\label{fig:18}
\end{center}
\end{figure}

The limit in Eq.~\eqref{eq:limit_ballistic} depends on the value of $p\in [0,1]$, as shown in Fig.~\ref{fig:17}.
On the other hand, the limits in Eqs.~\eqref{eq:limit_diffusive2} and \eqref{eq:limit_diffusive} do not depend on the value of $p$.
That fact is numerically confirmed in Fig.~\ref{fig:18}.

\clearpage

\section{Summary}
We studied an open quantum walk on $\mathbb{Z}$.
The walker changes its diffusion degree for time $t$ according to the values of parameters $\theta_0$ and $\theta_1$ which defines the one-step operations $P(\theta_0, \theta_1)$ and $Q(\theta_0,\theta_1)$, as shown in Table~1.
\begin{table}[h]
 \begin{center}
  \includegraphics[scale=0.4]{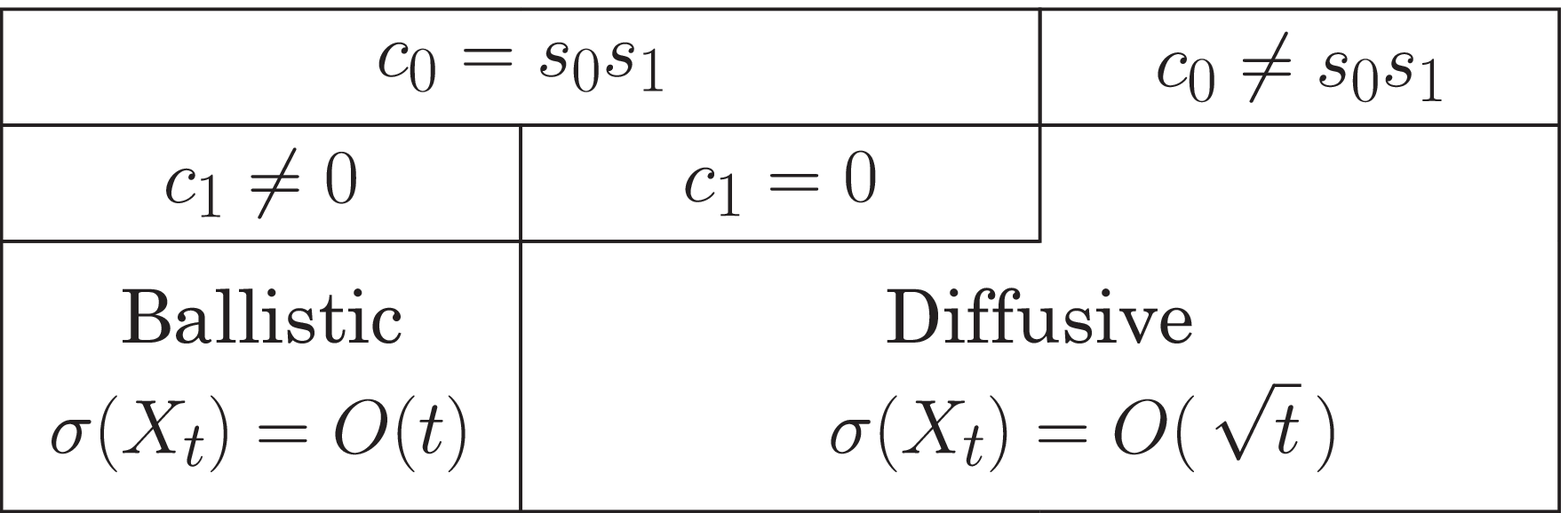}\\
  {\small Table~1. Phase transition of the open quantum walk}
 \end{center}
\end{table}

\begin{figure}[h]
 \begin{center}
  \includegraphics[scale=0.4]{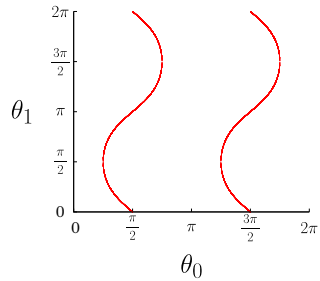}
  \caption{(color figure online) The set of $(\theta_0,\theta_1)\in [0,2\pi)\times [0,2\pi)$ such that $c_0=s_0s_1$.}
  \label{fig:19}
 \end{center}
\end{figure}

\begin{figure}[h]
\begin{center}
 \begin{minipage}{55mm}
  \begin{center}
   \includegraphics[scale=0.3]{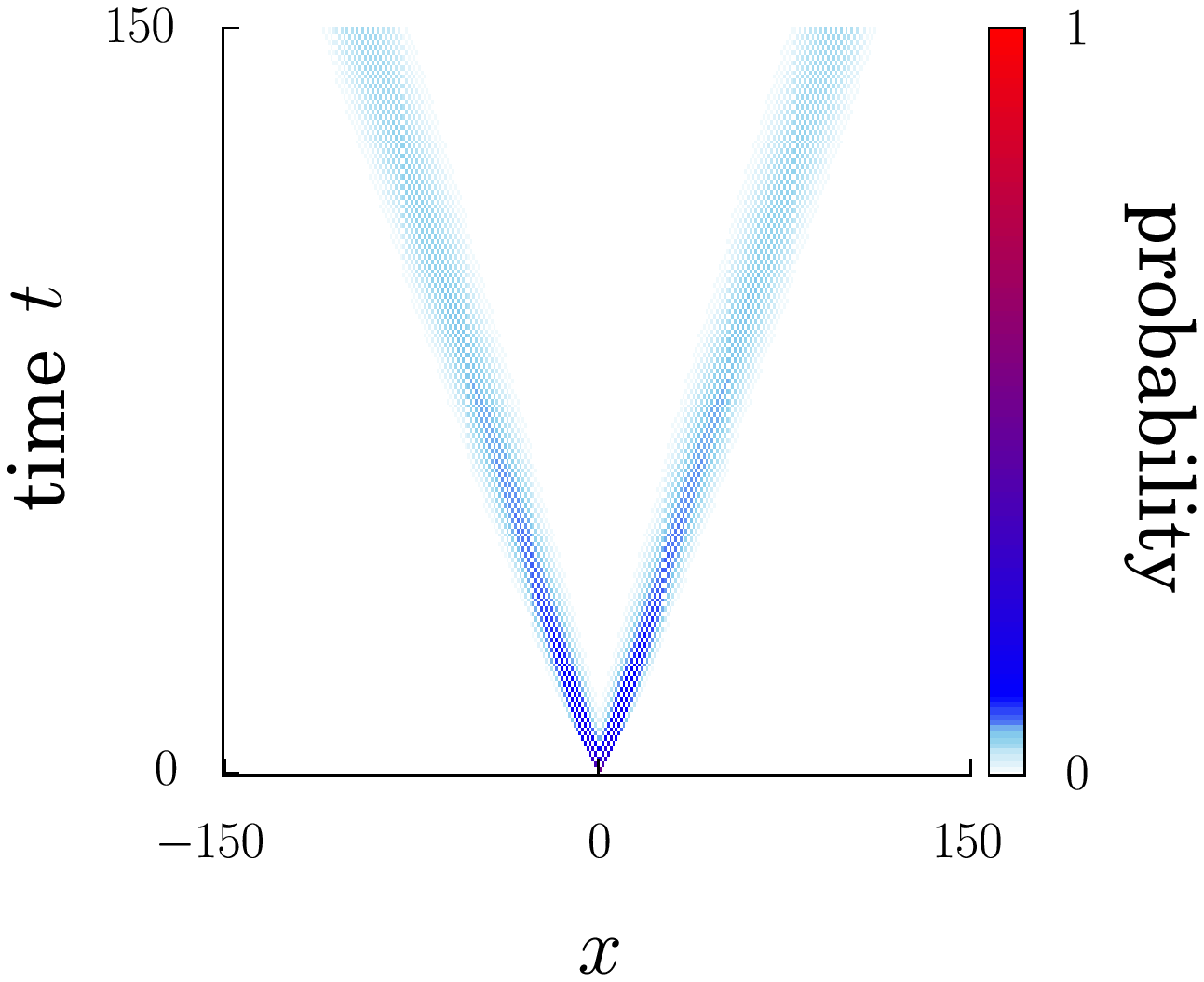}\\[2mm]
  (a) $\theta_0=2\pi/7,\,\theta_1=\arcsin(c_0/s_0)$
  \end{center}
 \end{minipage}
 \begin{minipage}{55mm}
  \begin{center}
   \includegraphics[scale=0.3]{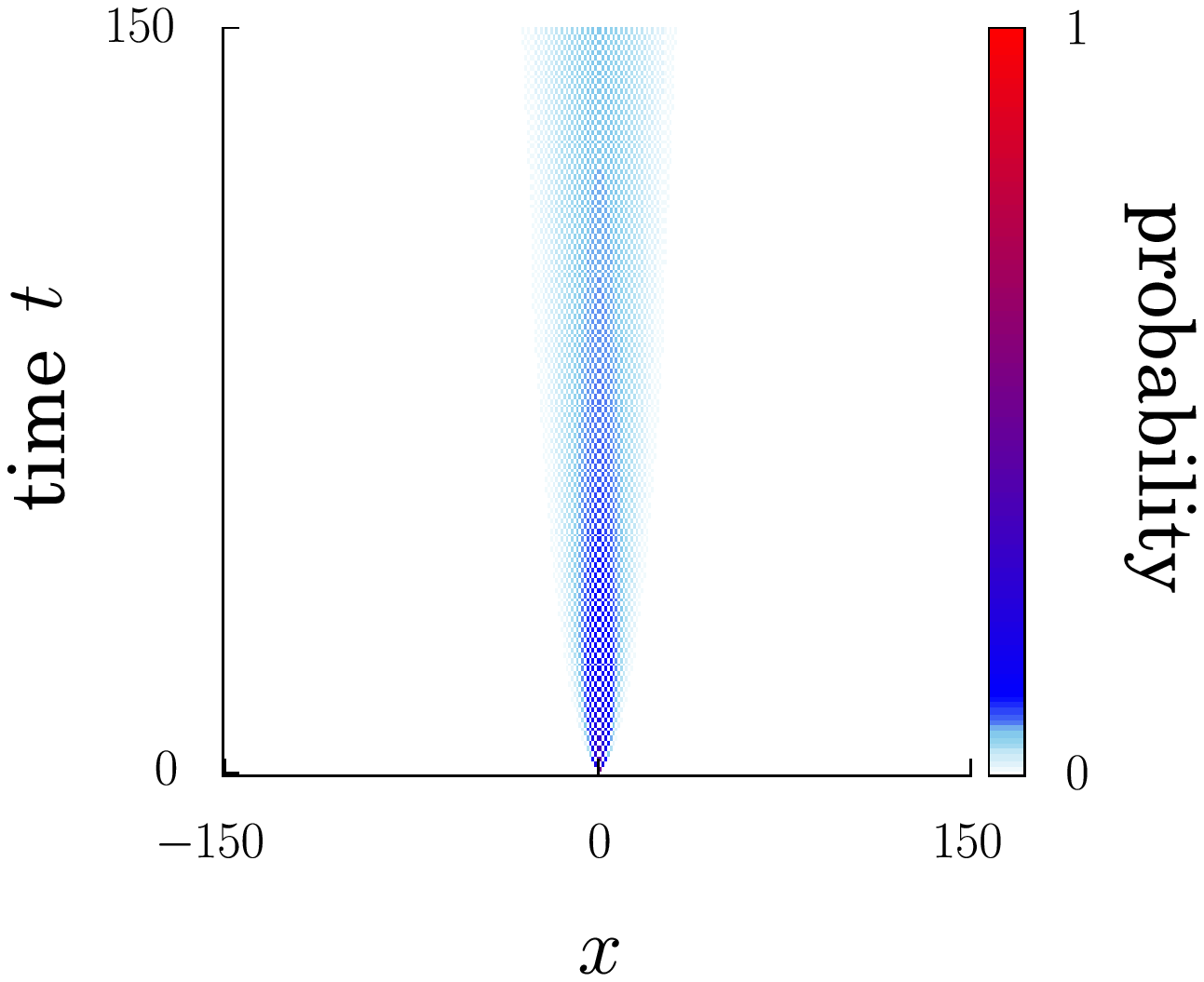}\\[2mm]
  (b) $\theta_0=\pi/6,\,\theta_1=\pi/3$
  \end{center}
 \end{minipage}
\vspace{5mm}
\caption{(color figure online) Time progress of the probability distribution $\mathbb{P}(X_t=x)$. The open quantum walker launches with the initial state $D_0=\ket{0}\otimes (1/2\ket{0}\bra{0}+1/2\ket{1}\bra{1})$.}
\label{fig:20}
\end{center}
\end{figure}

\noindent
The walker showed ballistic behavior if the parameters $\theta_0$ and $\theta_1$ satisfied the conditions $c_0=s_0s_1\,(\cos\theta_0=\sin\theta_0\sin\theta_1)$ and $c_1\neq 0\,(\cos\theta_1\neq 0)$, that is more precisely,
\begin{align}
 & \bigl\{(\theta_0, \theta_1)\, |\,  c_0=s_0s_1\, \mbox{and}\,  c_1\neq 0\bigr\}\nonumber\\[1mm]
 = & \bigl\{(\theta_0, \arcsin(c_0/s_0))\,|\, \theta_0\in (\pi/4, \pi/2] \cup (5\pi/4, 3\pi/2]\bigr\}\nonumber\\
 & \cup \{(\theta_0, \pi-\arcsin(c_0/s_0))\,|\, \theta_0\in (\pi/4, \pi/2] \cup (5\pi/4, 3\pi/2]\bigr\}\nonumber\\
 & \cup \bigl\{(\theta_0, \pi-\arcsin(c_0/s_0))\,|\, \theta_0\in (\pi/2, 3\pi/4) \cup (3\pi/2, 7\pi/4)\bigr\}\nonumber\\
 & \cup \bigl\{(\theta_0, 2\pi+\arcsin(c_0/s_0))\,|\, \theta_0\in (\pi/2, 3\pi/4) \cup (3\pi/2, 7\pi/4)\bigr\}.
\end{align}
Otherwise, the walker diffusively distributed.
Figure~\ref{fig:19} visualizes the set $\bigl\{(\theta_0, \theta_1)\, |\,  c_0=s_0s_1\bigr\}$.
A ballistic probability distribution and a diffusive probability distribution are numerically examined in Fig.~\ref{fig:20}.

We discovered a phase transition of an open quantum walk, but some details about the probability distribution $\mathbb{P}(X_t=x)$ still lack.
More precise analysis is needed and it would be a future challenge, for instance, a limit distribution as time $t\to\infty$.

\bigskip

The author is supported by JSPS Grant-in-Aid for Scientific Research (C) (No. 19K03625).
\bigskip


\end{document}